\documentclass{aa}  

\usepackage{graphicx}
\usepackage{hyperref}
\usepackage{txfonts}
\usepackage{orcidlink}
\usepackage{natbib}
\bibpunct{(}{)}{;}{a}{}{,} 
\usepackage{xcolor} 
\usepackage{booktabs} 
\usepackage{listings} 
\usepackage{csquotes} 
\usepackage[htt]{hyphenat} 

\definecolor{codegreen}{rgb}{0,0.6,0}
\definecolor{codegray}{rgb}{0.5,0.5,0.5}
\definecolor{codepurple}{rgb}{0.58,0,0.82}
\definecolor{backcolour}{rgb}{0.95,0.95,0.92}

\lstdefinestyle{mystyle}{
backgroundcolor=\color{backcolour},
commentstyle=\color{codegreen},
keywordstyle=\color{magenta},
numberstyle=\tiny\color{codegray},
stringstyle=\color{codepurple},
basicstyle=\ttfamily\footnotesize,
breakatwhitespace=false,
breaklines=true,
captionpos=b,
keepspaces=true,
numbers=left,
numbersep=5pt,
showspaces=false,
showstringspaces=false,
showtabs=false,
tabsize=2}
\usepackage{pifont}
\newcommand{\xmark}{\ding{55}}

\newcommand{\package}[1]{\textcolor{black}{\texttt{#1}}}

\newcommand{\ie}{i.e.\@}
\newcommand{\eg}{e.g.\@}
\newcommand{\cf}{cf.\@}

\newcommand{\kelvin}{\mathrm{K}}
\newcommand{\g}{\mathrm{g}}
\newcommand{\cm}{\mathrm{cm}}
\newcommand{\km}{\mathrm{km}}

\newcommand{\mpc}{\mathrm{Mpc}}

\newcommand{\s}{\mathrm{s}}

\newcommand{\ev}{\mathrm{eV}}
\newcommand{\kev}{\mathrm{keV}}
\newcommand{\mev}{\mathrm{MeV}}
\newcommand{\gev}{\mathrm{GeV}}
\newcommand{\erg}{\mathrm{erg}}

\newcommand{\MHz}{\mathrm{MHz}}
\newcommand{\GHz}{\mathrm{GHz}}
\newcommand{\Hz}{\mathrm{Hz}}
\newcommand{\rvir}{r_\mathrm{vir}}

\newcommand{\mvir}{M_\mathrm{vir}}

\newcommand{\kms}{\km\,\s^{-1}}
\newcommand{\cms}{\cm\,\s^{-1}}

\newcommand{\msol}{M_\odot}

\newcommand{\aqn}{AQN}
\newcommand{\aqns}{AQNs}
\newcommand{\qcd}{QCD}

\newcommand{\cp}{$\mathcal{CP}$}
\newcommand{\cs}{CS}
\newcommand{\maqn}{M_\mathrm{AQN}}
\newcommand{\naqn}{n_\mathrm{AQN}}

\newcommand{\raqn}{R_\mathrm{AQN}}

\newcommand{\ndw}{N_\mathrm{DW}}
\newcommand{\dw}{DW}
\newcommand{\seff}{\sigma_\mathrm{eff}}

\newcommand{\reff}{R_\mathrm{eff}}
\newcommand{\rcap}{R_\mathrm{cap}}
\newcommand{\microns}{\mu\mathrm{m}}

\newcommand{\rhogas}{\rho_\mathrm{gas}}
\newcommand{\ngas}{n_\mathrm{gas}}
\newcommand{\taqn}{T_\mathrm{AQN}}
\newcommand{\tgas}{T_\mathrm{gas}}
\newcommand{\ticm}{T_\mathrm{ICM}}

\newcommand{\dvel}{\Delta \mathrm{v}}
\newcommand{\diff}{\mathrm{d}}

\newcommand{\dfdnu}{\diff F(\nu)/\diff \nu}

\newcommand{\nneigh}{N_\mathrm{neigh}}
\newcommand{\weight}{\mathcal{W}}

\newcommand{\xrayemisint}{j_\mathrm{X\text{-}ray}}
\newcommand{\backgroundemis}{j_{\mathrm{background},\nu}}
\newcommand{\aqnemis}{j_{\mathrm{AQN},\nu}}

\newcommand{\ntemis}{j_{\mathrm{AQN, nonthermal},\nu}}
\newcommand{\xrayemis}{j_{\mathrm{X\text{-}ray},\nu}}

\newcommand{\emisunits}{\erg\,\cm^{-3}\Hz^{-1}\s^{-1}}
\newcommand{\emisintunits}{\erg\,\cm^{-3}\s^{-1}}
\newcommand{\samplea}{sample $\mathcal{A}$}
\newcommand{\sampleb}{sample $\mathcal{B}$}

\newcommand{\nutrans}{\nu_T}
\newcommand{\numin}{\nu_\mathrm{min}}

\newcommand{\pgas}{p_\mathrm{gas}}

\newcommand{\bbn}{BBN}
\newcommand{\tbbn}{T_\mathrm{BBN}}

\newcommand{\mqn}{MQN}
\newcommand{\mqns}{MQNs}

\newcommand{\slow}{\emph{SLOW}}

\newcommand{\specratio}{\tilde{\jmath}}

\begin{document} 

   \title{The Glow of Axion Quark Nugget Dark Matter: (II) Galaxy Clusters}

   \titlerunning{Axion Quark Nuggets in Simulated Galaxy Clusters}

   \author{
      Julian S. Sommer\inst{1}\,\orcidlink{0000-0002-1154-8317},
      Klaus Dolag\inst{1,2},
      Ludwig M. Böss \inst{1}\,\orcidlink{0000-0003-4690-2774},
      Ildar Khabibullin\inst{1,2}\,\orcidlink{0000-0003-3701-5882},
      Xunyu Liang\inst{3}\,\orcidlink{0000-0002-4230-6652},
      Ludovic Van Waerbeke\inst{3}\,\orcidlink{0000-0002-2637-8728},
      Ariel Zhitnitsky\inst{3}\,\orcidlink{0000-0002-2049-228X},
      Fereshteh Majidi\inst{3},
      Jenny G. Sorce\inst{4,5,6},
      Benjamin Seidel\inst{1},
      Elena Hernández-Martínez\inst{1}\,\orcidlink{0000-0002-1329-9246}
      }
\authorrunning{Sommer et al.}

   \institute{Universit\"ats-Sternwarte, Fakult\"at f\"ur  Physik, Ludwig-Maximilians Universität, Scheinerstr. 1, 81679 M\"unchen, Germany\\
   \email{jsommer@usm.uni-muenchen.de}
         \and
         Max-Planck-Institut für Astrophysik, Karl-Schwarzschild-Straße 1, 85741 Garching, Germany
         \and
         Department of Physics and Astronomy, University of British Columbia, Vancouver, V6T 1Z1, BC, Canada
         \and
         Univ. Lille, CNRS, Centrale Lille, UMR 9189 CRIStAL, 59000 Lille, France
          \and
          Université Paris-Saclay, CNRS, Institut d’Astrophysique Spatiale, 91405 Orsay, France\
          \and
          Leibniz-Institut f\"{u}r Astrophysik (AIP), An der Sternwarte 16, 14482 Potsdam, Germany
   }

   \date{Accepted XXX, Received YYY.}

 
  \abstract
   {The existence of axion quark nuggets is a potential consequence of the axion field, which provides a possible solution to the charge-conjugation parity violation in quantum chromodynamics. In addition to explaining the cosmological discrepancy of matter-antimatter asymmetry and a visible to dark matter ratio $\Omega_\mathrm{dark}/\Omega_\mathrm{visible}\simeq 5$, these composite compact objects are expected to represent a potential ubiquitous electromagnetic background radiation by interacting with ordinary baryonic matter. An in-depth analysis of axion quark nugget-baryonic matter interactions will be conducted in the environment of the intracluster medium in the constrained cosmological Simulation of the LOcal Web (SLOW).}
   {Here, we aim to provide upper limit predictions on electromagnetic counterparts of axion quark nuggets in the environment of galaxy clusters by inferring their thermal and non-thermal emission spectrum originating from axion quark nugget-cluster gas interactions.}
   {We analyze the emission of axion quark nuggets in a large sample of 161 simulated galaxy clusters using the SLOW simulation. These clusters are divided into a sub-sample of 150 galaxy clusters, ordered in five mass bins ranging from $0.8$ to $31.7 \times 10^{14} \,\msol$, along with 11 cross-identified galaxy clusters from observations. We investigate dark matter-baryonic matter interactions in galaxy clusters in their present stage at redshift $z=0$ by assuming all dark matter consists of axion quark nuggets. The resulting electromagnetic signatures are compared to thermal Bremsstrahlung and non-thermal cosmic ray synchrotron emission in each galaxy cluster. We further investigate individual frequency bands imitating the observable range of the WMAP, Planck, Euclid, and XRISM telescopes for the most promising cross-identified galaxy clusters hosting detectable signatures of axion quark nugget emission.}
   {We observe a positive excess in low and high energy frequency windows, where thermal and non-thermal axion quark nugget emission can significantly contribute to or even outshine the
   emission of the ICM in frequencies up to $\nu_T \lesssim 3842.19\,\GHz$ and $\nu_T \in [3.97,10.99]\times 10^{10}\,\GHz$, respectively. Emission signatures of axion quark nuggets are found to be observable if cosmic ray synchrotron emission of individual clusters is sufficiently low. The degeneracy of parameters contributing to an emission excess makes unambiguous predictions on pin-pointing specific regions of a positive axion quark nugget excess challenging, even though a general increase in the total galaxy cluster emission is expected by this dark matter model. Axion quark nuggets constitute an increment of $4.80\,\%$ of the total galaxy cluster emission in the low energy regime of $\nu_T \lesssim 3842.19\,\GHz$ for a selection of cross-identified galaxy clusters. We propose that the Fornax and Virgo clusters represent the most promising candidates to search for axion quark nugget emission signatures.}
   {The results from our simulations point towards the possibility of detecting an axion quark nugget excess in galaxy clusters in observations if their signatures can be sufficiently disentangled from the ICM radiation. While this model proposes a promising explanation for the composition of dark matter showing the potential to be verified by observations, we propose further changes to refine our methods to identify extracted electromagnetic counterparts of axion quark nuggets even with  more precision in the near future.}


   \maketitle

\section{Introduction}\label{sec:introduction}

During the Big Bang, a highly energetic environment sets the foundation for theories to emerge regarding matter and its composition. Evolved from different cosmological environments, modern theories describe a versatile zoo of dark matter models ranging from, for instance, sterile neutrinos \citep{Dodelson:1993je,Shi1999}, axion like particles \citep{Georgi:1986df,Masso:1995tw,Masso:2002ip}, weakly interacting massive particles \citep{Steigman:1984ac}, and quark nuggets that typically exhibit nuclear density \citep{Witten1984, Farhi1984, Rujula1984}. In addition to the comparatively low-mass dark matter candidates, massive astrophysical compact halo objects \citep{Alcock1993, Aubourg:1993wb} with for instance primordial mass black holes could propose a promising alternative as well \citep{Zeldovich:1967lct,Hawking:1971ei,Chapline:1975ojl}. Constraints on mass, properties, and number densities of these dark matter candidates can be made by various techniques. As for potential dark matter candidates, particle accelerators are instrumental in proposing mass boundaries for elementary particles, while gravitational wave detectors may aid in constraining the distribution of primordial black holes. A dark matter candidate is rarely observationally falsifiable by testing signatures in the electromagnetic spectrum.

Here we present a detailed analysis of the observational feasibility of a dark matter model that is proposed to obey properties of cold dark matter, but is still expected to be the origin of observable signatures that can be tested in a wide range of the electromagnetic spectrum, called the Axion Quark Nugget  \citep[\aqn,][]{Zhitnitsky2003}. 

\aqns{} are not only capable of describing generic cold dark matter features of structure formation but also provide solutions to fundamental cosmological problems. Two of the most prominent mysteries are: (i) Why is the abundance of dark matter not extraordinarily lower or larger than the visible component? (ii) Why do we observe more matter over antimatter, and where does this imbalance come from?

This paper is structured as follows: In \autoref{sec:overview}, a short overview will be provided by focusing on the formation, structure, and interaction scenario of AQNs in the respective subsections. Further, the foundation of interaction-resulting electromagnetic signatures will be set in \autoref{sec:methodology}, with a physical description responsible for the electromagnetic counterpart provided in \autoref{sec:taqn_heating} and \autoref{subsec:aqn_rad}. The implementation method using an SPH formalism in the constrained cosmological simulation \slow{} is presented in \autoref{sec:aqn_implementation}, with more details on the simulation provided in \autoref{sec:slow}. Selection criteria and details on our underlying sample are presented in \autoref{sec:sample}. Results will be provided in \autoref{sec:results}, followed by a discussion of our findings in \autoref{sec:discussion}. An outlook in conjunction with a concluding assessment is provided in \autoref{sec:conclusion}.

\section{A General Overview of AQNs}\label{sec:overview}

\subsection{The Formation Process}\label{sec:formation}

In this section, we describe the key steps for the formation of these particles and refer to \citep{Liang2016,Ge2017,Ge2018,Ge2019} for further details. Quantum Chromodynamics (\qcd) has a chiral anomaly term characterized by an axial angle $\theta$. This angle $\theta$ is physical and observable in the Standard Model. When $\theta\neq0$, the anomaly term induces a Charge-Conjugation Parity (\cp{}) violation, \ie{} the laws of physics do not behave in the same way under combined transformations of charge conjugation and parity. This leads to the so-called strong \cp{} problem, since the \cp{} violation in \qcd{} is not observed \citep{Abel:2020pzs}. The strong \cp{} problem can be naturally resolved when $\theta$ is considered a dynamical field so that the vacuum expectation value $\langle\theta\rangle$ settles to zero after the \qcd{} transition. Consequently, the dynamical field $\theta$ produces a hypothetical particle called axion\footnote{See original papers on the axion  \citep{Peccei1977, Weinberg1978, Wilczek1978, Kim1979, KSVZ2, Dine1981, DFSZ2} and recent reviews \citep{vanBibber:2006rb, Asztalos:2006kz, Sikivie:2008, Raffelt:2006cw, Sikivie:2009fv, Rosenberg:2015kxa, Marsh:2015xka, Irastorza:2018dyq, DiLuzio:2020wdo, Sikivie:2020zpn}.}. The Axion is very light and couples very weakly to ordinary matter.

In the framework of AQN, axions are produced before inflation. In this scenario, a unique physical vacuum occupies the entire Universe, and most types of topological defects [including the $N_{\rm DW}\neq1$ domain walls (\dw{}s)\footnote{Here $N_{\rm DW}$ refers to the number of physically distinct vacua.}] are forbidden. Only the $N_{\rm DW}=1$ \dw{}s, when $\theta$ interpolates in a single physical vacuum but different topological branches $\theta\rightarrow \theta + 2\pi n$ \citep{Vilenkin1982, Sikivie1982}, can be produced. The $N_{\rm DW}=1$ \dw{}s may form when the axion field starts to oscillate due to the misalignment mechanism at temperature $T_{\rm osc}\sim1{\rm\,GeV}$. This process continues until the QCD transition $T_{\rm c}\sim170{\rm\,MeV}$ when the quark chiral condensates form. From $T_{\rm osc}$ to $T_{\rm c}$, the axion mass effectively turns on, and the axion oscillation is underdamped. At the beginning of the oscillation $T_{\rm osc}$, the axion field is coherent globally. A coherent nonzero $\theta$ violates the global \cp{} symmetry and affects the \dw{} formation. Consequently, a preferred matter or antimatter species of nuggets will be formed -- it should be mentioned though, that an equal amount of nuggets would have been formed if $\theta=0$ at the moment of the \dw{} formation. Only the initial sign of the coherent $\theta$ field before oscillation dictates which nugget family is preferably formed. 

A small portion of the $N_{\rm DW}=1$ \dw{}s forms closed bubbles and acquires baryon charges from the quark-gluon plasma. Near and after the QCD transition $T_{\rm c}$, the \dw{}s on the bubbles can mix the axion with a tilted $\eta'$ field (the field of the $\eta'$ meson). Such a tilted substructure boosts the charge accumulation of the \dw{} bubbles. Depending on its inherent topological charge, a bubble acquires either matter or antimatter charges during this phase. Charge accumulation terminates at $T_{\rm form}\sim41{\rm\,MeV}$ and the bubbles form quark nuggets in the form of color superconducting (\cs{}) condensate. Due to the \cp{} violation effect discussed earlier, more antimatter AQNs are formed compared to matter AQNs. Under the assumption of zero baryon net charge in the Universe, the observed visible-to-dark-matter-density ratio of $\Omega_\mathrm{dark}:\Omega_\mathrm{visible}\simeq 5:1$ implies a baryon charge ratio of:
\begin{equation}\label{eq:aqn_distribution}
    B_\mathrm{\overline{AQN}}:B_\mathrm{AQN}:B_\mathrm{visible} \simeq 3:2:1\,,
\end{equation}
where the subscripts of $B$ correspond to the baryon charges of antimatter AQNs, matter AQNs, and visible matters respectively. Both, the matter- and the antimatter- AQNs serve as the DM component, while the remaining quarks in plasma become the visible component. Unlike conventional dark matter candidates such as weakly interacting massive particles (WIMPs) and axion, the AQN explains the observed cosmological relation $\Omega_\mathrm{dark}\sim\Omega_\mathrm{visible}$ without fine-tuning of fundamental parameters, since dark and visible matters now have the same QCD origin. In the AQN framework, freely propagating axions may exist via the misalignment mechanism \citep{Abbott1983, Dine1983, Preskill1983} and topological defects \citep{Chang1998, Kawasaki2015, Fleury2016, Klaer2017, Gorghetto2018}. However, they contribute a negligible amount to the dark sector $\Omega_{\rm dark}$ unless very specific fine-tuning applies \citep{Ge2018}. The conclusion \eqref{eq:aqn_distribution} is unaffected.

The \cs{} condensate in the AQN is stabilized by the excessive surface tension of the \dw{} bubble. The binding energy in the CS phase is sufficiently large (with a gap $\Delta\sim100{\rm\,MeV}$) so that AQNs do not modify the basic elements\footnote{In fact, it may resolve a long-standing primordial Lithium puzzle as argued in \cite{Flambaum:2018ohm}.} of the Big Bang Nucleosynthesis (\bbn) at $\tbbn\sim 1\,\mev$, where elements of metallicity $Z \leq 3$ were formed. Quark cores in the \cs{} state are in the lowest energy state possible, and it is assumed that quarks in neutron stars obey the same state \citep{Alford2008}. \aqns{} remain in this stable state over cosmic time scales.

On the other hand, if the theory of AQNs holds, the asymmetry of matter to antimatter in the observable Universe can be explained by an asymmetric baryon charge separation. A coherent background field $\theta$ violates the global \cp{} symmetry at the beginning of \dw{} bubble formation. This theory allows an equal amount of antimatter and matter, with more antimatter being hidden in composite nuggets than the ordinary one, causing it to be unobservable. These nuggets are stable over cosmic time, as the quark core is in a dense \cs{} state. It serves as cold dark matter because these macroscopic particles exhibit a low number density due to individual masses of the order of grams (a more specific description will be addressed in the following section). In summary, the AQN model provides intriguing cosmological implications by explaining the matter-antimatter asymmetry and the similarity between dark- and visible-matter densities. This is because dark and visible matters share the same QCD ancestor.

\begin{figure*}
    \centering
    \includegraphics[width=\linewidth]{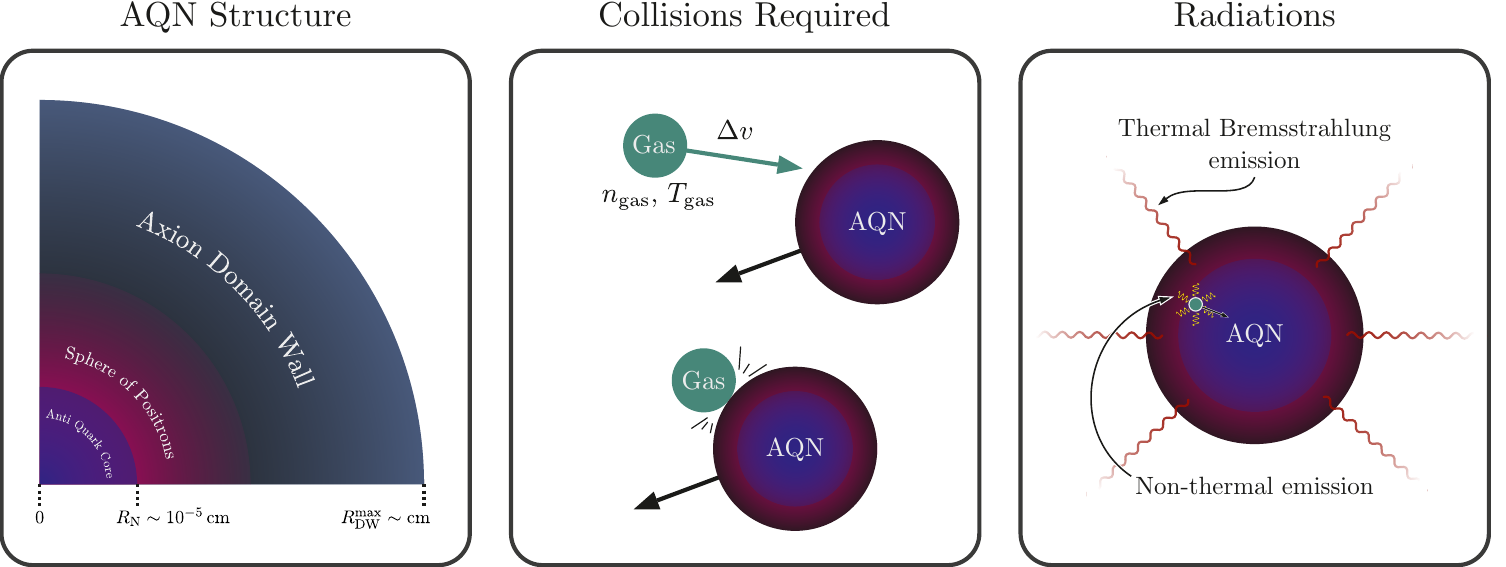}
    \caption{Schematic sketch for the antimatter-\aqn's internal structure (not true to scale) with an indication of typical sizes (left panel) and the processes, which lead to the different radiation signatures (middle and right panel). The antimatter-\aqn{} consists of a core of anti-quarks surrounded by positrons. When colliding with the ambient gas, the antimatter-\aqn{} heats up and radiation occurs, which depends on the speed at which the collision occurs, the gas temperature and density as well as the ionization state of the gas. The right panel shows the two key emissions, namely thermal and non-thermal emissions. From an incoming proton (green, small circle), $2\,\gev$ of energy will be available after collision with the \aqn. Through annihilation, energetic quarks and gluons are produced and stream either deep within the nugget (around 90\% of the $2\,\gev$) or toward the nugget surface (around 10\% of the $2\,\gev$). Quarks and gluons, which remain inside, will contribute to the thermalization of the nugget. Heat will be transferred from the core to the electrosphere which cools down via thermal bremsstrahlung emission. Quarks and gluons, which proceed to the surface without thermalization, can transfer energy to positrons and produce non-thermal emission.}
    \label{fig:aqn_schematic}
\end{figure*}

\subsection{Structure of AQNs and Cosmological Properties}\label{sec:aqn_struct}

It is important to differentiate matter families as (anti-)\aqn{} consist of a core of (anti)quarks surrounded by positrons/electrons that make the electrosphere. The electrosphere is surrounded by the $\ndw = 1$ axion \dw{}. Interactions of baryonic matter with the axion \dw{} seldom occur. The surface tension of the \dw{} exerts a strong pressure on the quark core causing it to preserve in a \cs{} state. To prevent the \dw{} from collapsing further, Fermi pressure from the \aqn{} counteracts the \dw{} surface tension.

Parameters that describe the properties of an AQN are the baryon number $B$, the electric charge $eQ$, where $Q$ typically corresponds to the number of positrons that depleted from an originally neutral \aqn{} and in some scenarios the magnetization $\mathcal{M}$ \citep{Santillan2020}, which will be neglected in this study for reasons of simplification. It will be shown in later sections that $Q$ highly depends on the temperature of the surrounding gas as positrons can deplete if the conditions suffice. The lower limit on $B$ can be observationally constrained by non-detections from IceCube with $\langle B\rangle > 3\times 10^{24}$ \citep{Lawson2019}. Similar limits are also obtainable from the ANITA and geothermal data \citep{Gorham:2012hy}. $B$ results in the \aqn's mass $\maqn$ and size $\raqn$, with most recent discussion on their distribution in \cite{Raza2018,Ge2019}. Since mass contributions from the electrosphere are negligible, the baryonic mass component of an \aqn{} can be estimated as follows:

\begin{equation}\label{eq:maqn}
    \maqn \approx m_p B \simeq 16.7 \left(\frac{B}{10^{25}}\right)\,\g\,.
\end{equation}

In the state of a \cs{} phase, one can assume the nuclear density to be $\rho_n = 3.5\times 10^{14}\,\g\,\cm^{-3}$ \citep{Zhitnitsky2018} resulting in an \aqn{} radius of:

\begin{equation}
    \raqn = \left(\frac{3\maqn}{4\pi\rho_n}\right)^{1/3} \approx 2.25\times 10^{-5}\,\cm.
\end{equation}

Ionized \aqns{} are characterized by a charge $Q$, with $Q$ being a function of $\taqn$. A small fraction of positrons in the electrosphere are loosely bound and in the non-relativistic Boltzmann regime \citep{Forbes2008b}, which rises with the internal \aqns{} temperature. 

Throughout this paper, we will fix the parameters, physically describing the structure of \aqns{} to $B = 10^{25}$, $\maqn = 16.7\,\g$ and $\raqn = 2.25\times 10^{-5}\,\cm$. A distribution function of these parameters is not well-constrained yet and would drastically increase the complexity of the system.

\subsection{Possible Interaction Scenarios}\label{sec:interactions}

As pointed out in \autoref{sec:aqn_struct}, \aqns{} consist of two different particle families and will therefore yield different combinations of possible interaction scenarios. Not all of the different interactions are likely to be visible, and therefore their relevance for our study will be assessed in the following:

\begin{enumerate}
    \item \textbf{Antimatter \aqn{} interaction with electrons:} Electrons can interact with the electrosphere made up of positrons for antimatter \aqns. These $e^+e^-$ annihilations might possibly explain the $511\,\kev$ line emission coming from the galactic center \citep{Oaknin2005a, Zhitnitsky2007, Forbes2010a, Flambaum:2021xub} and the COMPTEL gamma-ray photons in an energy range of $1-20\,\mev$ \citep{Lawson2008}.
    \item \textbf{Antimatter \aqn{} interaction with protons:} Free protons propagating through the electrosphere can collide with the antiquark core (see \autoref{fig:aqn_schematic}). Each collision produces an annihilation energy of $2\,\gev$. A major fraction (about 90\%) of the annihilation energy will thermalize the quark core and transfer to the electrosphere. This heating process leads to the emission of the electrosphere in the form of thermal Bremsstrahlung. Observations, that can be explained by thermal \aqn{} emission are for instance the diffuse galactic microwave excess \citep{Forbes2008b}, the $21\,\cm$ absorption line \citep{Lawson2019a}, and diffuse galactic UV background emission \citep{Zhitnitsky2022a}. A minor fraction (about 10\%) of annihilation energy will be emitted as a short pulse of non-thermal photons at the point of impact. Observations, like Chandra's diffuse $k_B T\approx 8\,\kev$ emission \citep{Forbes2008a} might find their nature in non-thermal \aqn{} emission processes.
    \item \textbf{Antimatter \aqn{} annihilation with celestial bodies:} As it is discussed in detail by \cite{Liang2022}, antimatter \aqns{} can annihilate with more matter at the same time. Axions, residing off-shell in the domain wall can be emitted, when \aqns{} lose potential energy after annihilation processes in the core took place. To retain a minimum total energy of the system and therefore maintain the stability of the \aqn, axions will be emitted to lower the \dw's mass. In addition to axions, antimatter AQNs produce electromagnetic radiation as described in (1) and (2). As a consequence, phenomena like for instance coronal heating, Extreme Ultraviolet (EUV) and X-ray emissions might be attributed to the outcomes of \aqn{} interactions in the sun \citep{Zhitnitsky2017, Zhitnitsky2018, Raza2018, Ge2020}. The nature of fast radio bursts may be explained by \aqns{} by triggering magnetic reconnection events in magnetars as well \citep{Waerbeke2019}. Furthermore, according to \cite{Budker2022}, infrasound, acoustic and seismic waves might be the direct outcome of \aqn-earth encounters. Non-inverted polarity detections by ANtarctic Impulse Transient Antenna (ANITA) were discussed by \cite{Liang2022b}, and short time bursts detected by the Telescope Array (TA) were analyzed by \cite{Zhitnitsky2021, Liang2022a} in the framework of \aqn{} annihilations. Cosmic ray-like events detected by the Pierre Auger Observatory might be explained by \aqns{} inducing lightning strikes with subsequent direct emission \citep{Zhitnitsky2022}. Multi-modal events detected by Horizon-T that were discussed using \aqns{} by \cite{Zhitnitsky2021a} should be distinct from conventional cosmic-ray showers.
    \item \textbf{Matter \aqn{} interaction with antimatter:} Technically, one could expect the same emission processes from (1) and (2) with matter \aqns{} with positrons and antiprotons. However, because of the low abundance of antimatter cosmic rays (\eg{} \citep{Hillas2006, Blum2017}) and the slightly lower abundance of matter \aqns{} over antimatter \aqns, it is expected that matter \aqns{} remain dark for most of the time and electromagnetic signals originating from matter \aqns{} can be neglected.
    \item \textbf{Matter \aqn-Antimatter \aqn{} interaction:} Even though the electromagnetic outcome of matter-antimatter \aqn{} interactions would result in an impressive electromagnetic signature, it is because of the low number density of $\naqn\sim 10^{-29}\,\cm^{-3}$ that this interaction scenario can be entirely neglected (see comments on this for example in \cite{Zhitnitsky2022}). And although both of the \aqns{} can have an attractive influence on each other if ionized, the higher effective cross-section would not drastically increase interaction rates.
    \item \textbf{\aqn-\aqn{} interaction from same particle family:} Since \aqns{} from the same particle family would only collide by chance with no interaction enhancement by their ionization state, collisions are expected to be even rarer than in (5). An electromagnetic signature would at most be visible by thermalization of a heated \aqn{} after collision.
\end{enumerate}

As we will assume the intracluster medium (ICM) to consist of ionized hydrogen, this work will only focus on emission phenomena coming from proton-antimatter \aqn{} interactions. Line emission from $e^+e^-$ annihilation, axion emission by strong \aqn{} emission, and \aqn-\aqn{} self-interactions will be neglected in this study as thermal and non-thermal \aqn{} emission from proton interactions is presumed to dominate the spectrum.

With the fixed parameters describing a single \aqn, one can infer the antimatter \aqn's number density by assuming the mass ratio relation from \autoref{eq:aqn_distribution}. According to the mass ratio relation, a fraction of $3/5$ of the entire dark matter accounts for anti-\aqns. Since ionized gas in the ICM mainly consists of normal matter, the highest rate of annihilation processes -- and hence visible interactions -- will be expected from antimatter \aqns. One third of the total \aqn{} mass comes from the $\ndw=1$ axion \dw{} \citep{Ge2018}. Therefore -- to account for the antimatter \aqn{} number density $\naqn$ from the total \enquote{observable} dark matter density $\rho_\mathrm{DM}$ -- we would further have to multiply a factor of $2/3$ by excluding the \dw{} contribution to the AQN mass. We therefore obtain an anti-\aqn{} number density estimate of:

\begin{equation}\label{eq:naqn}
    \naqn = \frac{2}{3}\times\frac{3}{5}\times\frac{\rho_\mathrm{DM}}{\maqn},
\end{equation}

\noindent
which typically scales in a galaxy cluster environment as $\naqn\sim 10^{-29}\,\cm^{-3}$. It is because of the low number density that \aqns{} mostly preserve cold dark matter features -- provided they are in a suitable environment, however, the occurring interactions can even yield different emission processes of \aqns. For the sake of simplicity, we will further refer to antimatter axion quark nuggets when using the abbreviation \enquote{AQN}.

\section{Methodology}\label{sec:methodology}

\subsection{Internal AQN Temperatures Through Ambient Gas Interactions}\label{sec:taqn_heating}

The internal temperature of \aqns{} is influenced by annihilation processes of protons impacting a nugget with anti-quarks from the inner core. The internal temperature of a nugget can be estimated from the conservation of energy. In thermal equilibrium, if a proton is captured by an AQN, the injected energy must be equal to the radiative output. Protons can be captured more efficiently if the anti-quark nugget is ionized, but loses efficiency under certain conditions of the environment -- we, therefore, have to introduce an effective cross-section $\seff$ to model collision rates. The energy conversion efficiency during a collision depends on a proxy for the probability of quantum reflection for neutral gas called $(1-f)$, and the fraction of thermal photon production $(1-g)\approx 0.9$. Because of the increased Coulomb potential in an ionized gas, the quantum reflection probability $\mathcal{P}=(1-f)$ is highly suppressed, and we therefore obtain $f\sim 1$. The Coulomb potential causes the hydrogen ion to be trapped around an AQN until the final collision and therefore annihilation occurs. The injected energy per unit time can be estimated by:

\begin{equation}
    \frac{\diff E}{\diff t} = (1-g)f \seff \ngas\Delta E\dvel \,.
\end{equation}

$\Delta E$ is the energy that will be released by an annihilation event of a single proton amounting up to $\Delta E\sim 2m_p c^2\approx 2\,\gev$ and $\dvel=|v_\mathrm{AQN}-v_\mathrm{gas}|$ is the relative AQN-proton speed. The radiative output is the luminosity of a nugget with radius $\raqn$, \ie:

\begin{equation}\label{eq:laqn}
    L=4\pi \raqn^2F^\mathrm{th}\,.
\end{equation}

If we set $\diff E/\diff t=L$ and after re-expressing the effective cross-section $\seff=\kappa\pi\raqn^2$ using the scaling parameter $\kappa$, we obtain the expression:

\begin{equation}\label{eq:energy_flux}
    F^\mathrm{th}=2\,\gev\,\frac{\kappa}{4}(1-g)f\ngas\dvel \,.
\end{equation}

As a consequence of the internal temperature increase due to the annihilation process, the positrons in the electrosphere will be heated and emit Bremsstrahlung. In natural units, \ie{} $c=k_\mathrm{B}=\hbar=1$ and $h=2\pi$, the total surface emissivity of an \aqn{} is calculated by \cite{Forbes2008b} and is given by:

\begin{equation}\label{eq:aqn_bremsstrahlung}
    F^\mathrm{th}\approx\frac{16\alpha^{5/2}}{3\pi}\taqn^4\left(\frac{\taqn}{m_e}\right)^{1/4}.
\end{equation}

$\alpha$ is the fine-structure constant and $\taqn$ is the internal temperature of the AQN. After equating \autoref{eq:energy_flux} with \autoref{eq:aqn_bremsstrahlung}, one can solve for $\taqn$ and obtains: 

\begin{equation}\label{eq:taqn_1}
    \taqn = \left(\frac{3\pi}{4} \frac{2\, \gev}{16\alpha^{5/2}}(1-g)f\kappa\, m_e^{1/4} \ngas\dvel\right)^{4/17}.
\end{equation}

\autoref{eq:taqn_1} can be represented more intuitively if we use galaxy cluster-typical gas densities of $n_\mathrm{gas,\,cluster}\sim 10^{-3}\cm^{-3}$ and a relative velocity of $\dvel\sim 10^{-3} c$:

\begin{multline}\label{eq:taqn_2}
    \taqn \approx 6.95\times 10^{-2}\,\ev \left(\frac{\ngas}{10^{-3}\,\cm^{-3}}\right)^{4/17} \left(\frac{\dvel}{10^{-3}\,c}\right)^{4/17} \\
    \times \left(\frac{(1-g)f}{0.9}\right)^{4/17}\kappa^{4/17}.
\end{multline}

The numerical factor $\kappa$ boosts the internal \aqn{} temperature if the effective cross-section is larger than its geometrical cross-section. We therefore use the following condition for each \aqn:

\begin{equation}\label{eq:kappa_cases}
    \kappa=
  \begin{cases}
    \left(\frac{\reff}{\raqn}\right)^2, & \reff > \raqn \\
    1, & \reff \leq \raqn
  \end{cases}
\end{equation}

To find an expression for $\reff$, two important contributions have to be taken into account. First, the charge of an \aqn{} $Q$, which is a proxy for the ionization stage, and second, the temperature of the surrounding plasma. In galaxy clusters, the ICM consists of highly ionized gas with temperatures of $\ticm\sim\kev$. In an entirely ionized gas, \cite{Zhitnitsky2023} shows that the cross-section of \aqns{} can scale with $\reff$ instead of $\raqn$, leading to a potential increase in the collision efficiency with the surrounding gas. It is important to note that geometrical radii of \aqns{} are of the order of $\raqn \sim 0.1\,\microns$ and -- given that the gas number density in galaxy clusters is of the order of $\ngas\sim 10^{-3}\,\cm^{-3}$ -- collision rates with surrounding particles can be quite low if $\reff < \raqn$. It is not sensible to incorporate effective radii smaller than their geometrical size, and therefore for $\reff < \raqn$ we will set $\reff=\raqn$. It is possible to derive $\reff$ from the physical properties of the nugget and its environment. 

In the following \cite{Zhitnitsky2023} proposes that the potential energy of attraction for \aqns{} in an ionized gas scales with the thermal energy of the plasma in the environment, \ie:

\begin{equation}\label{eq:thermal_pot}
    \frac{\alpha Q}{\rcap}\sim \frac{m_p v_p^2}{2}\sim \tgas\,,
\end{equation}

\noindent
with the fine-structure constant $\alpha$, the charge $Q$, the proton mass $m_p$ and proton velocity $v_p$. \autoref{eq:thermal_pot} shows at which radius an ionized \aqn{} can capture protons from the gas with thermal velocities of $\tgas\sim m_p v_p^2$. What follows from this scaling relation is a capture radius which decreases for increasing gas temperatures. \cite{Zhitnitsky2023} used the capture radius instead of the effective radius with $\rcap = \sqrt{\pi} \reff$. In this paper, however, we will adapt the convention of using $\reff$ instead of $\rcap$. After re-expressing \autoref{eq:thermal_pot}, we obtain the following relation:

\begin{equation}\label{eq:reff_1}
    \reff = \frac{\alpha Q}{\tgas\sqrt{\pi}}\,.
\end{equation}

It is important to note that $Q$ is a function of $\taqn$. The higher $\taqn$, the higher the thermal motion of positrons in the electrosphere with positrons following a non-relativistic Boltzmann regime \citep{Forbes2008b}. If kinetic positron energies exceed the potential energy of the \aqn, positrons can deplete, leading to a negative charge for \aqns. The positrons are rather weakly bound and the number of positrons $Q$ which are likely to evaporate from the AQN can be estimated by using the mean field approximation \citep{Forbes2008b} of the local density of positrons $n(z,\taqn)$ at a distance $z$ from the surface of an AQN. According to \cite{Zhitnitsky2023}, one obtains for typical $\taqn$ in cluster environments:

\begin{align}
    Q &\simeq \frac{4\pi \raqn^2}{\sqrt{2\pi\alpha}}(m_e\taqn)\left(\frac{\taqn}{m_e}\right)^{1/4} \label{eq:q} \\
    &\approx 4.62\times 10^3 \left(\frac{\taqn}{10^{-2}\,\ev}\right)^{5/4}\left(\frac{\raqn}{2.25\times 10^{-5}\,\cm}\right)^2, \label{eq:ionisation}
\end{align}

\noindent
with $m_e$ and units of $\cm$ from \autoref{eq:q} being re-expressed in units of $\ev$ using natural units (\ie{} $c=\hbar=1$ and $h=2\pi$). Now, after using \autoref{eq:taqn_1}, \autoref{eq:reff_1}, \autoref{eq:ionisation} and $\kappa = (\reff/\raqn)^2$, one finds:

\begin{multline}\label{eq:reff_full}
    \reff \simeq 3.89\times 10^{-7}\,\cm\left(\frac{\ngas}{10^{-3}\cm^{-3}}\right)^{5/7}\left(\frac{\dvel}{10^{-3}c}\right)^{5/7} \\
    \times \left(\frac{\raqn}{2.25\times 10^{-5}\cm}\right)^{24/7}\left(\frac{1\,\kev}{\tgas}\right)^{17/7}.
\end{multline}

\autoref{eq:reff_full} immediately implies that in a galaxy cluster environment, the majority of AQNs will be treated with $\kappa=1$, since $\raqn>\reff$.

\subsection{Radiation Processes of AQNs}\label{subsec:aqn_rad}

\aqns{} can emit both thermal and non-thermal radiation. The major fraction $(1-g)\approx0.9$ of the annihilation energy of $2\,\gev$ is emitted by thermal Bremsstrahlung. The remaining fraction $g\approx0.1$ is radiated non-thermally. The total surface emissivity was already introduced in \autoref{eq:aqn_bremsstrahlung} which was derived by \cite{Forbes2008b}:

\begin{equation}\label{eq:flux_from_emissivity}
    F_\nu^\mathrm{th} \sim \frac{8}{45}\alpha^{5/2}\taqn ^3\left(\frac{\taqn}{m_e}\right)^{1/4}\times f(x),
\end{equation}

\noindent
with $f(x)$ being:

\begin{equation}\label{eq:cases}
  f(x)=(1+x)e^{-x}
  \begin{cases}
    (17-12\ln(x/2)), & x < 1 \\
    (17+12\ln(2)), & x \geq 1,
  \end{cases}
\end{equation}

\noindent
and $x=2\pi\nu/\taqn$. \cite{Majidi2023} rewrote \autoref{eq:flux_from_emissivity} in physical units. This also changes $x$ to $\tilde{x}=2\pi\hbar\nu/(k_B\taqn)$, such that we get:

\begin{equation}\label{eq:thermal_spec_nat}
    F_\nu^\mathrm{th} = \frac{8}{45}\frac{\alpha^{5/2}k_B^3 \taqn ^3}{\hbar^2 c^2}\left(\frac{k_B \taqn}{m_e c^2}\right)^{1/4}\times f(\tilde{x}).
\end{equation}

\begin{figure}
    \centering
    \includegraphics[width=\linewidth]{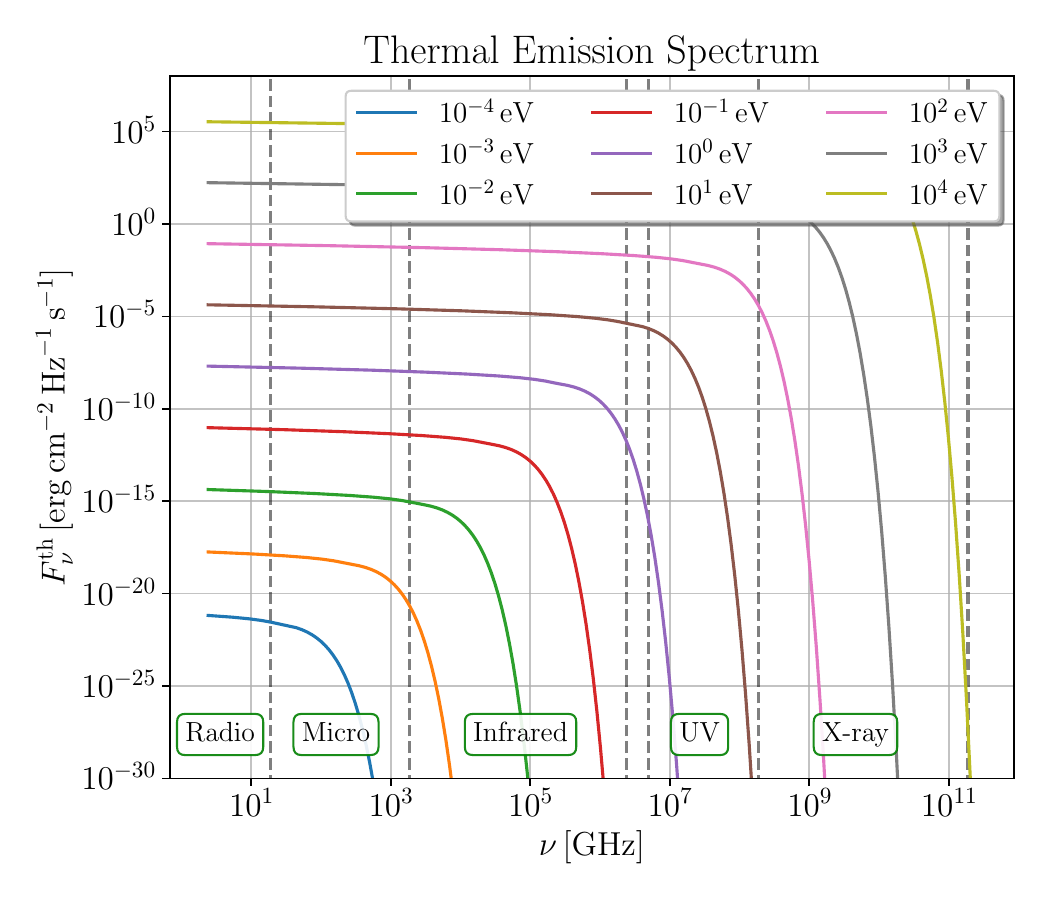}
    \caption{Spectral range for thermal emission from \aqns. The magnitude and width of a thermal spectrum increase for increasing $\taqn$ and can reach from radio to X-ray frequencies for extraordinary hot \aqns.}
    \label{fig:thermal_spectrum}
\end{figure}

\autoref{fig:thermal_spectrum} shows a plot of $\dfdnu$ for a set of internal AQN temperatures for $\nu\in 2.42\times[10^8, 10^{20}]\,\Hz$. Temperatures that fall into the cluster regime are mostly abundant in the radio- and microwave bands.

In the case of non-thermal emission, \cite{Forbes2008a} proposed the non-thermal emission per frequency to scale with: 

\begin{equation}
    F_\nu^\mathrm{non\text{-}th} = \frac{\nu}{\nu_c}\int_{\nu/\nu_c}^\infty K_{3/5}(x)\,\diff x,
\end{equation}

\noindent
with $K_{3/5}(x)$ being the second modified Bessel function. The critical frequency $\nu_c = \omega_c/(2\pi)\approx 30\,\kev/(2\pi)$ was approximated by \cite{Forbes2008a}. It is important to note that $\nu_c=30\,\kev/(2\pi)$, whose numerical value is set to 30 for conventional reasons, is not strictly constrained. For the expression of the synchrotron function, various approximations can be estimated by different approaches and integration methods (see for example \cite{Crusius1986, Rybicki1986, Weniger1990, Seidel2003, Fouka2013, Fouka2014, Yang2017, Palade2023}). For our study, we will adapt the approximation of the integrated modified Bessel function from \cite{Majidi2023}:

\begin{equation}\label{eq:majidi_synch}
    F_\nu^\mathrm{non\text{-}th} \approx C(\beta) x^\beta e^{-x},
\end{equation}

\noindent
with $C(1/3)\approx 1.81$. It is important to note that $\beta$ is usually not fixed and can vary depending on the energy distribution of positrons. Additionally, the exponent depends on the emission mechanism at the nugget's surface, which can be free-free-like emission rather than synchrotron-like, resulting in $\beta=0$. Given the various analytical approximations and the choice of utilizing the synchrotron function to model the non-thermal spectrum requiring more physical evidence, one has to keep in mind that the non-thermal \aqn{} spectrum is not constrained well enough to propose definite predictions. With this in mind, the non-thermal \aqn{} emission has to be understood with caution. Nevertheless, the bolometric surface emissivity of the non-thermal component can then be calculated by substituting $x$ with $\nu/\nu_c$ in \autoref{eq:majidi_synch} and integrating over $\nu\in[0,\infty]$. The final expression of $F_\nu^\mathrm{non\text{-}th}$ is derived in great detail in \cite{Majidi2023} and will be directly adapted from the mentioned literature:

\begin{equation}\label{eq:majidi_nt_reff}
    F_\nu^\mathrm{non\text{-}th} = \frac{2\gev fg}{4\Gamma(4/3)}\frac{\ngas\dvel}{\nu_c}\left(\frac{\reff}{\raqn}\right)^2\left(\frac{\nu}{\nu_c}\right)^{1/3}e^{-\nu/\nu_c},
\end{equation}

\noindent
and expressed in terms of $\taqn$:

\begin{equation}\label{eq:majidi_nt_taqn}
F_\nu^\mathrm{non\text{-}th} = \frac{16 \alpha^{5/2}}{3 \pi\, \Gamma(4/3)}\frac{k_B^{17/4}}{\hbar^3 c^{5/2}}\frac{g}{1-g}\frac{1}{\nu_c}\frac{T_{\rm AQN}^{17/4}}{m_e^{1/4}} \left(\frac{\nu}{\nu_c}\right)^{1/3}e^{-\nu/\nu_c}\,.
\end{equation}

\begin{figure}
    \centering
    \includegraphics[width=\linewidth]{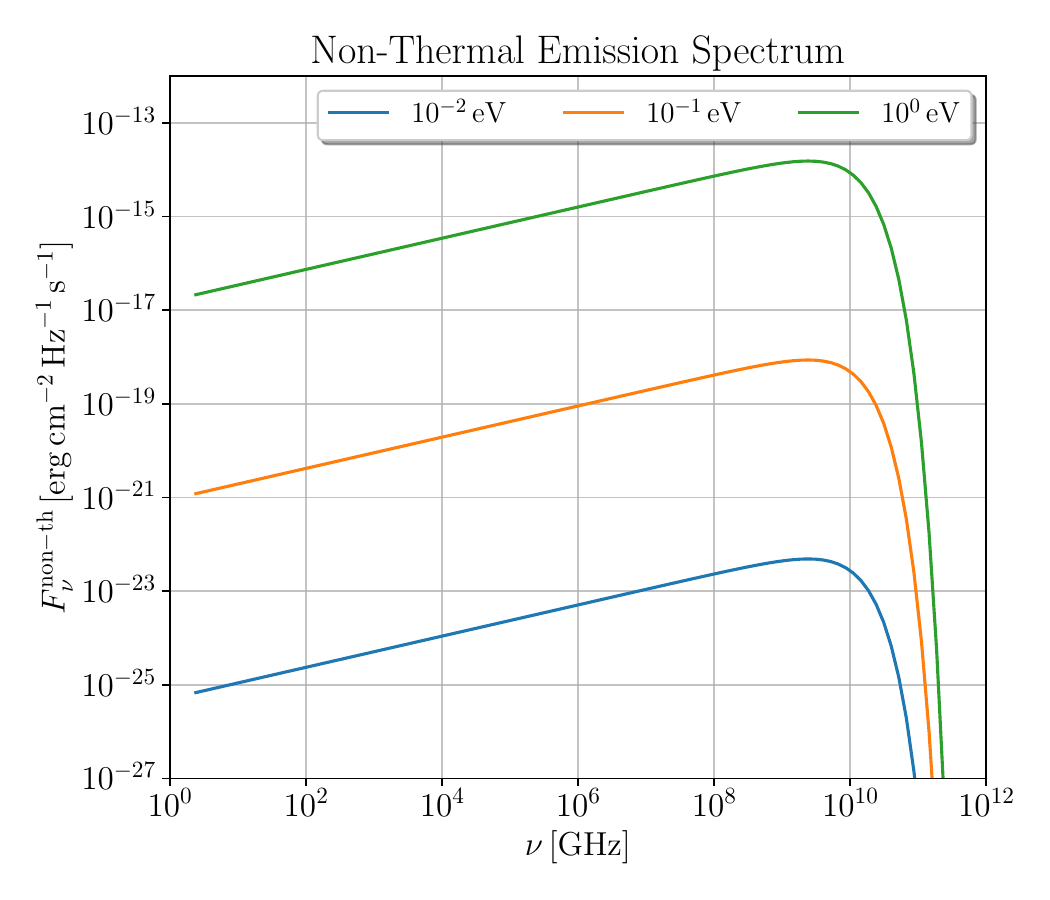}
    \caption{Non-thermal \aqn{} emission spectrum for different $\taqn$ serving as a validation to reproduce Figure 3 from \cite{Majidi2023}.}
    \label{fig:nonthermal_spectrum}
\end{figure}

A plot of the non-thermal spectrum for typical $\reff$, $\ngas=10^{-3}\cm^{-3}$ and $\dvel=10^7\cms$ in a galaxy clusters is shown in \autoref{fig:nonthermal_spectrum}. In most cases, $\reff=\raqn$, and when comparing this to \autoref{fig:thermal_spectrum} for typical \aqn{} temperatures in galaxy clusters ($\taqn \in [10^{-2},10^{-1}]\,\ev$), it can be seen that the non-thermal emission is expected to be a few orders of magnitude lower than the thermal emission.

In a system of multiple \aqns, a more appropriate representation of radiation is the emissivity $j_\nu$, which depends on the number density of \aqns{} $\naqn$, and it is therefore crucial to evaluate individual local number densities instead of assuming a global $\naqn$ (the denser the environment, the more emission we can expect). It is therefore important to evaluate the emissivity of each \aqn{} tracer particle individually. With the individual emissivity, one can obtain a cluster spectrum or emission maps after summing over all individual $j_\nu$. A more detailed description of how local properties in a particle simulation are inferred will be presented in \autoref{sec:aqn_implementation}. We will adapt \autoref{eq:naqn} for describing the number density, while we have to keep in mind that $\rho_\mathrm{DM,i}$ should be evaluated at the $i$th \aqn{} tracer particle in the simulation. The emissivity is represented by:

\begin{equation}\label{eq:aqn_emissivity}
    j_{\nu,\,i} = 4\pi R_\mathrm{AQN}^2 n_{\mathrm{AQN},\,i} F_{\nu,\,i}.
\end{equation}

\subsection{Mapping AQN Properties onto Halo Tracer Particles}\label{sec:aqn_implementation}

Cosmological simulations in the Smoothed-Particle Hydrodynamics (SPH) formalism use tracer particles to represent environmental properties in a specific region. Depending on the resolution in our simulation, a tracer has a mass of $M_\mathrm{part} \sim 10^9-10^{10}h^{-1} \msol$. In the underlying \aqn{} model, a particle represents properties of temperature, density, and velocity relative to the environment within the region it exists in. In our analysis of the simulation, particle families are split into gas and halo tracer particles. The motion of halo tracer particles is defined only by their gravitational interactions with their surroundings. It is because of this property that halo tracer particles serve as tracers for \aqns{} in this analysis. As proposed in \autoref{eq:taqn_2}, physical properties of \aqns{} rely on the cluster environment -- more specifically, on the underlying gas properties of the ICM. It is not trivial that gas properties of SPH tracer particles can be mapped onto a single halo tracer particle, and it will therefore be described in more detail how the mapping process is numerically implemented.

Initially, the number of neighbouring gas particles that can influence the \aqn{} properties is defined to be $\nneigh = 200$. To reduce the computational costs, we use the package \package{NearestNeighbors.jl} \citep{Carlsson2022} and sort particles by the $k$-d tree algorithm to find the nearest gas neighbors surrounding an \aqn{} particle.

Now that for each \aqn{} all gas particles are localized, the relative velocities in each of the $x$, $y$, and $z$-coordinates to all neighbouring gas particles are calculated to infer the speed, which is then saved in an array. Depending on their distance relative to the gas particles, \aqns{} are characterized by different weights $\weight$ for each neighbouring gas particle. Kernel weights are usually a function of the corresponding kernel function $k_K$, the \aqn{} distance to the $j$th gas particle $d_{i,j}$, and the smoothing length $h_{\mathrm{AQN},i} = \max(d)$ if $d \neq 0$ of an \aqn{} with respect to the $j$th gas particle, \ie{} $\weight_\mathrm{\aqn}(k_K, d_{i,j}, h_{\mathrm{AQN},i})$, and can be calculated by using the package \package{SPHKernels.jl} \citep{Boess2023b}. A 3-dimensional Wendland C2 kernel is chosen for \aqns{} and a 3-dimensional Wendland C4 for gas particles (see \cite{Dehnen2012} for choosing kernels). When considering the $i$th \aqn{} particle, any arbitrary SPH property $A_j$ of the $j$th gas particle can be calculated according to \cite{Dolag2008} via:

\begin{equation}\label{eq:aqn_sph_dv}
    \langle A_i\rangle = \sum_{j=1}^{\nneigh} \frac{m_{\mathrm{gas},j}}{\rho_{\mathrm{gas},j}}A_j\weight_\mathrm{\aqn}(k_\mathrm{C2}, d_{i,j}, h_{\mathrm{AQN},i}).
\end{equation}

In this case, this is how $\langle\dvel_i\rangle$ is calculated. However, to properly infer gas properties at positions of \aqn{} particles relative to their environment, it is crucial to define a kernel that computes weights only from the given gas \texttt{HSML} property and their corresponding distances to the \aqn. For the gas properties, we therefore obtain the mapped gas particle SPH property onto the \aqn{} via:

\begin{equation}\label{eq:aqn_sph_rest}
    \langle A_i\rangle = \sum_{j=1}^{\nneigh} \frac{m_{\mathrm{gas},j}}{\rho_{\mathrm{gas},j}}A_j\weight_\mathrm{\aqn}(k_\mathrm{C4}, d_{i,j}, h_{\mathrm{gas},j}).
\end{equation}

This finally gives us the following set of equations that have to be solved numerically:

\begin{equation}
    \langle \Delta \mathrm{v}_i\rangle = \sum_{j=1}^{\nneigh} \frac{m_{\mathrm{gas},j}}{\rho_{\mathrm{gas},j}} \Delta \mathrm{v}_j \weight_\mathrm{\aqn}(k_\mathrm{C2}, d_{i,j}, h_{\mathrm{AQN},i})
\end{equation}

\begin{equation}\label{eq:number_density_sph}
    \langle n_{\mathrm{gas},i} \rangle = \sum_{j=1}^{\nneigh} \frac{m_{\mathrm{gas},j}}{\rho_{\mathrm{gas},j}} n_{\mathrm{gas},j} \weight_\mathrm{\aqn}(k_\mathrm{C4}, d_{i,j}, h_{\mathrm{gas},j})
\end{equation}

\begin{equation}
    \langle T_{\mathrm{gas},i} \rangle = \sum_{j=1}^{\nneigh} \frac{m_{\mathrm{gas},j}}{\rho_{\mathrm{gas},j}} T_{\mathrm{gas},j} \weight_\mathrm{\aqn}(k_\mathrm{C4}, d_{i,j}, h_{\mathrm{gas},j}).
\end{equation}

Here, $h_{\mathrm{gas},j}$ can be directly read from the \texttt{HSML} block of each gas particle. To ensure that numerical values do not fall below sensible SPH values, the conditions

\begin{align}
    \langle T_{\mathrm{gas},i}\rangle & =
    \min(\tgas)\text{ for }\langle T_{\mathrm{gas},i}\rangle < \min(\tgas) \\
    \langle n_{\mathrm{gas},i}\rangle & =
    \min(\ngas)\text{ for }\langle n_{\mathrm{gas},i}\rangle < \min(\ngas)
\end{align}

\noindent
are applied. For reasons of simplification, only the absolute relative velocity for each particle was taken to calculate an SPH representative for the \aqn, which can be calculated via:

\begin{equation}
    \Delta \mathrm{v}_{ij} = \sqrt{(v_{j,\mathrm{gas}}^{\,x}-v_{i,\mathrm{AQN}}^{\,x})^2+(v_{j,\mathrm{gas}}^{\,y}-v_{i,\mathrm{AQN}}^{\,y})^2+(v_{j,\mathrm{gas}}^{\,z}-v_{i,\mathrm{AQN}}^{\,z})^2}.
\end{equation}

It is a direct consequence of this approach that $\langle \Delta \mathrm{v}_i\rangle$ is obtained independently of the direction of the velocity vector of a gas particle relative to the \aqn. Only a single scalar is saved to describe the smoothed speed of an \aqn{} with respect to its environment. It is therefore possible that particles are strongly weighted because of their relative distance to the \aqn{} particle but have already passed the \aqn{} and are moving away. This circumstance, on the other hand, is not too big of an issue, since tracer particles are representations of particle distributions, and direct collisions are therefore not desired in the setup of an SPH simulation.

\subsection{Additional Radiation Sources in Simulated Galaxy Clusters}

\subsubsection{ICM Bremsstrahlung}

It is important to compare the \aqn{} emission to gas emission to infer \aqn{} excesses and to identify potential emission windows. Amongst other sources of radiation in the X-ray continuum, Bremsstrahlung coming from the hot ICM is the most dominant emission process (\cite{Felten1969}, and \cite{Sarazin1986} for an overview). Following \autoref{sec:aqn_implementation}, two different SPH gas properties are accessible: the gas properties at each \aqn{} position and the gas properties at each gas particle. For gas emission, the latter will be chosen, since the spatial distribution of gas particles differs from the ones of halo tracer particles. It is specifically important to analyze different regions where gas emission is more prominent and where \aqn{} positions might differ. The X-ray emissivity $\xrayemis$ in the energy band $[\nu_0,\nu_1]$ depends on the hydrogen mass fraction $X_H$, the gas temperature $\tgas$, the gas density $\rhogas$, on the gaunt factor $g(\nu,\tgas)$ and the frequency $\nu$. It has the following form:

\begin{multline}\label{eq:xray_emissivity}
    \xrayemisint = 4 C_j g(\nu,\tgas)(1+X_H)^2\left(\frac{\tilde{n}}{\tilde{f}m_p}\right)^2\rhogas^2[\g\,\cm^{-3}]\sqrt{\tgas\,[\kev]} \\
    \times \left(\exp\left(-\frac{h\nu_0\,[\kev]}{\tgas\,[\kev]}\right)-\exp\left(-\frac{h\nu_1\,[\kev]}{\tgas\,[\kev]}\right)\right)\,\emisintunits,
\end{multline}

\noindent
with

\begin{equation}
    \tilde{n} = \frac{X_H+0.5(1-X_H)}{2X_H+0.75(1-X_H)},
\end{equation}

\noindent
and

\begin{equation}
    \tilde{f} = \frac{4}{5X_H+3},
\end{equation}

\noindent
where $X_H=0.76$ is a parameter of the simulation, and $C_j=2.42\times 10^{-24}$ is a numerical factor provided by \cite{Bartelmann1996}. Since a detailed comparison of the \aqn{} and gas spectra requires the emissivity per unit frequency and \autoref{eq:xray_emissivity} only provides the emissivity in an integrated form, we will have to take the derivative per frequency bin to obtain the units $\emisunits$. In the numerical implementation, a frequency array of 1000 elements was defined in the frequency range of $\nu\in[10^{-5},10^{8}]/(2\pi)\,\ev$ given in log-space. Each consecutive element has an increased factor of $a=\nu_{i+1}/\nu_i$, which is constant for any $i$. For the factor $a$, we can therefore choose $i=0$ for simplification, \ie{} the first element in the frequency array. $\Delta \nu_i = a\nu_i-\nu_i$ is then the $i$th frequency band, providing a sufficient approximation for the derivation with:

\begin{equation}
    \xrayemis(\nu) = \frac{\xrayemisint(\nu)}{\Delta\nu_i}
\end{equation}

\subsubsection{Estimating the Synchrotron Emission of the ICM}\label{sec:crs}

There are various sources of synchrotron emission within galaxy clusters. For a detailed, recent review on radio emission from galaxy clusters, see \citep{vanWeeren2019review}. Most importantly for our work, some clusters show large-scale volume filling radio emission which roughly follows the distribution of the ICM, so-called giant radio halos. Their exact nature is still under debate and theoretical models involve turbulent re-acceleration processes and/or secondary electrons from hadronic interactions. Other sources like peripheral radio emissions, so-called relics that are related to shocks or radio emission related to individual galaxies, or active galactic nuclei and their jets will be neglected for this first study. Given the large uncertainty involved when directly modeling the radio emission from first principles from our simulation (see discussion in \cite{boess2023slow}), we decided to follow to more conservative approach to produce the possible synchrotron signal from the ICM when comparing to the predicted AQN signal. Here we made use of the fact that observations of a large sample of galaxy clusters suggest that, compared to the Coma galaxy cluster, other clusters typically show less or even no diffuse radio emission \citep{Hanisch1982radio, Cuciti2021haloes}. In addition, due to the quasi-near linear correlation observed between the radio signal and the Sunyaev-Zeldovich effect \citep{Planck2013Coma} we can directly relate the monochromatic radio emissivity to the pressure of the ICM. The radio halo of Coma is also observed at various frequencies so that an overall spectrum can be used \citep{Thierbach2003coma}. Making use of having a very good counterpart of Coma in our constrained simulation \citep{Dolag2023, Hernandez2024} we, therefore, assume a scaling of the pressure within the simulation with the radio emissivity at $352\,\MHz$, which reproduces the scaling presented in \cite{Planck2013Coma} and modified at different frequencies to reproduce the spectrum measured in \cite{Thierbach2003coma}. This very well reproduces the observed scaling relation between mass and radio power at $1.4\,\GHz$ \citep{Cuciti2021haloes}, although with this approach every cluster shows a giant radio halo and therefore gives an upper limit of the expected synchrotron emission from the ICM in the simulations.

\subsection{The Constrained Simulation of the Local Universe \enquote{SLOW}}\label{sec:slow}

The underlying simulation is a constrained cosmological simulation of the local universe, called \enquote{Simulating the LOcal Web (SLOW)} \citep{Dolag2023}. Derived from the CosmicFlows-2 catalog \citep{Tully2013}, the peculiar velocity field can help to constrain initial conditions for a simulation of our local universe, when tracing back the trajectories up to a defined redshift. Tracers, like velocity and density fields, are direct proxies for the underlying gravitational fields which are important for the initial conditions of the simulation. This study uses the constrained cosmological simulated simulation, covering a volume of $(500\,h^{-1}\mpc)^3$ and including $3072^3$ gas and dark matter particles. It assumes a cosmology based on Planck measurements \citep{Planck2014}, with a Hubble constant $H_0 = 67.77\kms\,\mpc^{-1}$, a total matter density $\Omega_M=0.307115$, a cosmological constant $\Omega_\Lambda = 0.692885$, a baryon fraction $\Omega_B = 0.0480217$, a normalization of the power spectrum $\sigma_8 = 0.829$ and a slope of the primordial fluctuation spectra $n = 0.961$ as described in \cite{Dolag2023}.

\subsection{The Sample}\label{sec:sample}

In this work, we tested the spectral emission of \aqn{} quantitatively for a large sample and qualitatively for a small sample. A large sample of 150 galaxy clusters (which we call \enquote{\samplea{}}) ordered in 5 corresponding mass bins was used to test scaling relations and mass dependencies for the search of emission excesses in the spectra. Each mass bin was defined to contain 30 galaxy clusters with bin sizes of $M_{\mathrm{vir},1}\in [0.8,0.9]\times 10^{14}\msol$, $M_{\mathrm{vir},2}\in [1.1,2.0]\times 10^{14}\msol$, $M_{\mathrm{vir},1}\in [4.0,5.0]\times 10^{14}\msol$, $M_{\mathrm{vir},1}\in [7.0,7.9]\times 10^{14}\msol$ and $M_{\mathrm{vir},1}\in [10.7,31.7]\times 10^{14}\msol$. Mass bins were arbitrarily chosen to maintain a relatively equal mass distribution and conserve a well-distributed population of virial masses. The lowest mass range was chosen to be $0.8\times 10^{14}\msol$ since it becomes tricky at lower masses to separate galaxy clusters from galaxy groups. As an upper mass range, the $31.7\times 10^{14}\msol$ were obtained by taking the most massive galaxy clusters above the boundary of $10^{14}\msol$. We used the subfind algorithm \citep{Springel2001, Dolag2009} implemented into the \package{GadgetIO.jl} package \citep{Boess2023c} to receive the necessary halo IDs to read out the halo and gas tracer particles within a certain box.

\enquote{Sample $\mathcal{B}$} consists of 11 out of 45 cross-identified galaxy clusters \citep{Hernandez2024} that resemble digital twins to their real counterparts, which was then utilized to identify the most promising galaxy cluster candidates that exhibit \aqn{} signatures. According to \cite{Hernandez2024}, galaxy clusters from observations were cross-identified with simulated galaxy clusters by comparing their mass, X-ray luminosity, temperature, and Compton-$y$ signal. A probability of cross-identification can be established by measuring the probability of obtaining each cross-match in a random, unconstrained simulation. By computing the significance, the authors could assess how well the cross-identified clusters are associated with each other. The most massive galaxy cluster from the digital twin-sample is Coma ($\mvir = 18.82\times 10^{14}\,\msol$) and the least massive is Fornax ($\mvir = 0.61 \times 10^{14}\,\msol$).

Galaxy clusters from both \samplea{} and \sampleb{} were read out from a sphere of radius $r=b \times \rvir$, where $b=1.5$ is a chosen multiplier used to extend the sphere.

\section{Results}\label{sec:results}

\subsection{Environmental Impact onto Physical Properties of AQNs}

\begin{figure*}
    \centering
    \includegraphics[width=\linewidth]{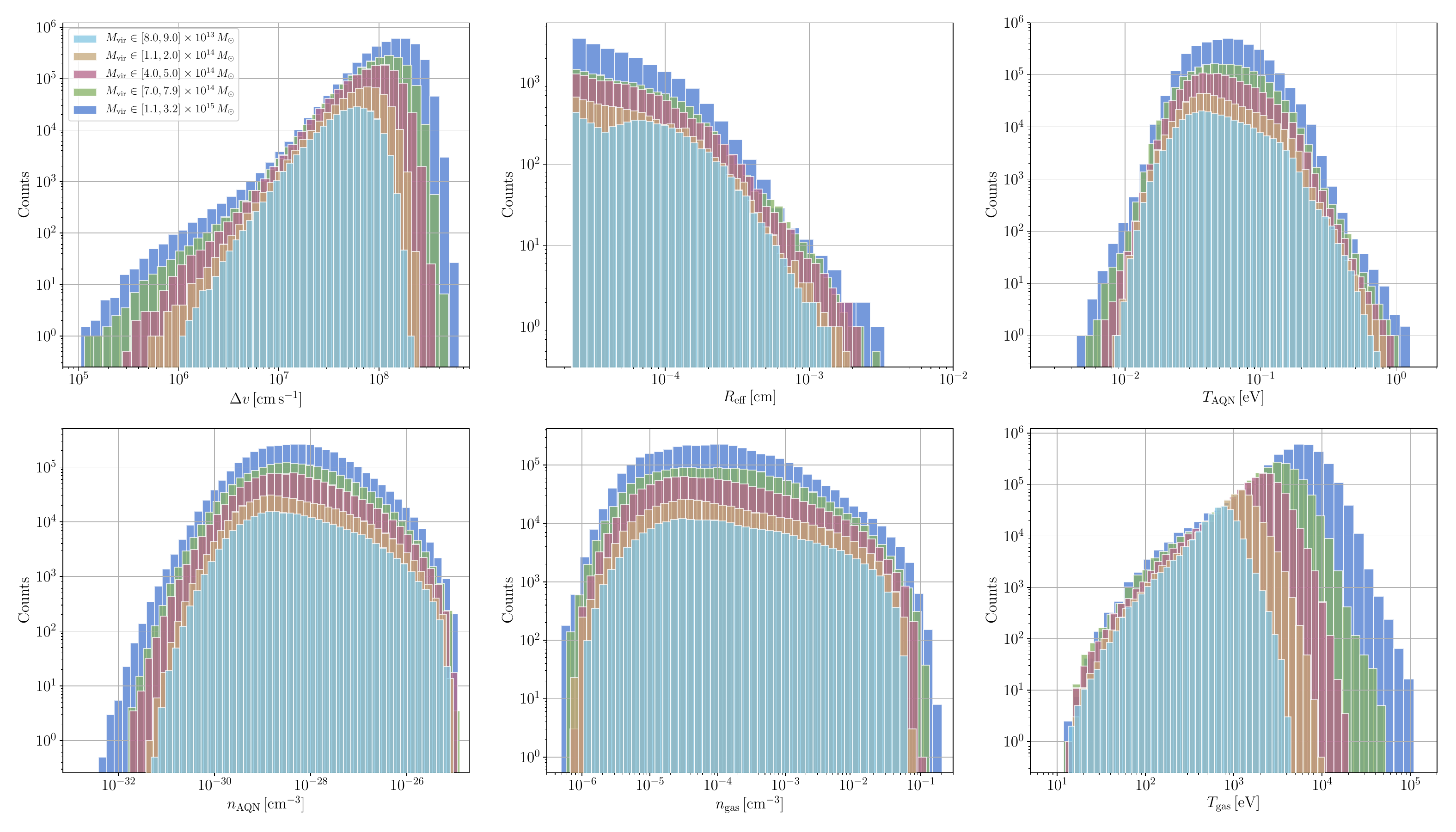}
    \caption{Histograms of of physical properties relevant to the \aqn{} emission. Median values were taken for each mass bin, \ie{} for 30 galaxy clusters, respectively.}
    \label{fig:histograms}
\end{figure*}

Histograms in \autoref{fig:histograms} show the distribution of $\taqn$, $\dvel$, $\naqn$, $\reff$, $\ngas$ and $\tgas$, respectively. Out of the distributions of all 150 galaxy clusters, the median distribution in each mass bin, \ie{} for 30 galaxy clusters each, was taken and plotted. It can be seen that -- the larger the range gets -- all properties show a relatively wide distribution with fewer outliers. Since the cluster mass directly scales with the particle numbers, a general trend of increased total counts can be seen in all histograms -- however it is interesting to see how the shape of the histograms changes for increasing mass bins.

\begin{itemize}
    \item \textbf{Relative velocity:} In the $\dvel$-histogram, most of the particles are distributed towards $\dvel\sim 10^8\cms\sim 0.3\% c$. Intriguingly, an additional low-velocity population component of $\dvel\lesssim 10^6\cms$ seems to be present, especially in massive galaxy clusters. A $\dvel$ peak shift towards higher velocities for larger cluster masses can also be observed. 
    \item \textbf{Internal \aqn{} temperature:} The $\taqn$-histogram is heavily weighted for values of $\taqn\in[10^{-2},10^{-1}]\,\ev$ and occasionally high $\taqn\gtrsim 1\,\ev$ can be possible for massive clusters. Peaks in the distribution shift from approximately $4\times 10^{-2}\,\ev$ to $6\times 10^{-2}\,\ev$ for increasing cluster masses. Here, $\dvel$ is likely to play the dominating factor (as discussed before). On the low $\taqn$-side, fewer counts are caused by non-optimal combinations of $\dvel$, $\ngas$ and $\tgas$. The higher $\taqn$-side is limited by an increasing amount of high counts of high $\tgas$.
    \item \textbf{Effective \aqn{} radius:} The $\reff$-histogram directly shows that most of the $\reff$ are set to $\raqn = 2.25\times 10^{-5}\,\cm$, since $\kappa=1$ in most environments. Depending on the cluster mass, the general shape of the $\reff$ distribution does not drastically change.
    \item \textbf{AQN number density:} Following \autoref{eq:naqn}, values in the $\naqn$-histogram are generally low, since \aqns{} are large and massive compared to elementary particles. Therefore, a smaller abundance of \aqn{} is required, which results in low $\naqn$.
    \item \textbf{Gas number density:} $\ngas$ shows a reasonable distribution with a heavier weight in number densities of $\ngas < 10^{-3}\cm^{-3}$ because of the extended readout-radius of $1.5\,\rvir$.
    \item \textbf{Gas temperature:} For increasing $\mvir$ in galaxy clusters the peak in the $\tgas$-histograms shifts to larger values. Higher gas temperatures for heavier galaxy clusters can be expected due to a steeper potential that in return attracts more substructures. A constant infall of new galaxies continuously heats the ICM and provides a general increase in $\tgas$ (see \cite{Sarazin1986}, for a review).
\end{itemize}

\begin{figure*}
    \centering
    \includegraphics[width=\linewidth]{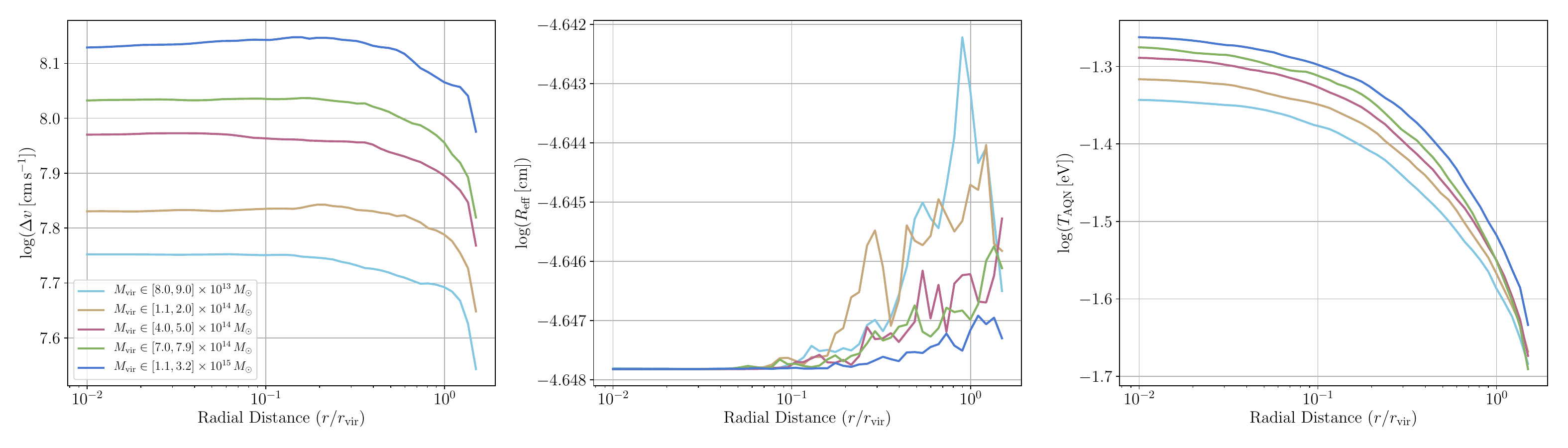}
    \caption{Radial profiles of $\dvel$, $\reff$, and $\taqn$ mapped onto \aqn{} tracer particles. The median radial profile was taken for each mass bin (\ie{} 30 galaxy clusters each) and compared to other mass bins in a region of $r\in[0,1.5]\,\rvir$.}
    \label{fig:rad_profs}
\end{figure*}

A better understanding of the spatial distribution of \aqn{} properties in galaxy clusters can be obtained when considering their radial profiles. \autoref{fig:rad_profs} shows how $\dvel$, $\taqn$, and $\reff$, mapped onto the position of \aqns{} scale with increasing distances from the cluster's center of mass. Radial profiles were taken from each galaxy cluster in \samplea{} and the median radial profile was taken for each mass bin. SPH properties were inferred according to the proposed methodology in \autoref{sec:aqn_implementation}.

\begin{itemize}
    \item \textbf{Relative velocity:} An interesting outcome of the radial $\dvel$ profiles is that for the most massive galaxy clusters, the profile increases towards normalized distances of $r/\rvir\sim 0.2$ and decreases at larger distances. All mass bins show relatively flat radial profiles until $r/\rvir\sim 0.1$.
    \item \textbf{Effective \aqn{} radius:} $\reff$ represents an intriguing property of \aqns{} since (due to the high ICM temperature in the central regions) the gas properties do not permit effective radii to be larger than $\raqn$. At larger distances, individual regional properties enable small increases in $\reff$ that highly depend on the degenerate \aqn{} properties present in the surrounding gas. Single substructures might be capable of obtaining large effective radii, and it is subject to \autoref{sec:cluster_maps} to identify these regions in spatially resolved maps of individual galaxy clusters to properly infer their physical properties. It is important to note that a slight trend of increasing $\reff$ can already be observed at $r \leq 1.5\,\rvir$. Even though the increase is almost negligible with a maximum increase of $\reff$ hosted by the lowest cluster mass bin of $\Delta\reff\simeq 3 \times 10^{-7}\,\cm$ (\ie{} an increase of $\sim 1.3\%$), the phenomenon of higher $\reff$ in the peripheries becomes more significant the further the distance is. This is a feature of the underlying \aqn{} model that was anticipated for this paper, since \aqns{} are expected to be surrounded by a fully ionized gas. In regions of $r\gg 1.5\,\rvir$, smaller $\tgas$ values consequently yield large $\reff$, as ionized particles can be captured over larger distances if the thermal motion of the ions is sufficiently low. This will lead to a runaway effect for increasing $\reff$ that is not physically meaningful. It is, therefore, crucial to constrain the sample to galaxy cluster regions with an extent of approximately $r_\mathrm{max}\sim1.5\,\rvir$ at maximum for the underlying \aqn{} model.
    \item \textbf{Internal \aqn{} temperature:} $\taqn$ are most likely increasing for different masses with higher $\dvel$. For larger radii and in all mass bins, $\taqn$ seems to converge towards a single radially normalized value. Given the underlying \aqn{} model, it is assumed that $\taqn$ will artificially rise for radii $r\gg 1.5\,\rvir$, because $\reff\rightarrow\infty$ for $\tgas\rightarrow 0\,\kelvin$. Moreover, this means that at a certain relative distance, $\taqn$ will mainly be influenced by the properties of filaments and voids.
\end{itemize}

\subsection{Spectral Energy Distributions}

\begin{figure*}
    \centering
    \includegraphics[width=\linewidth]{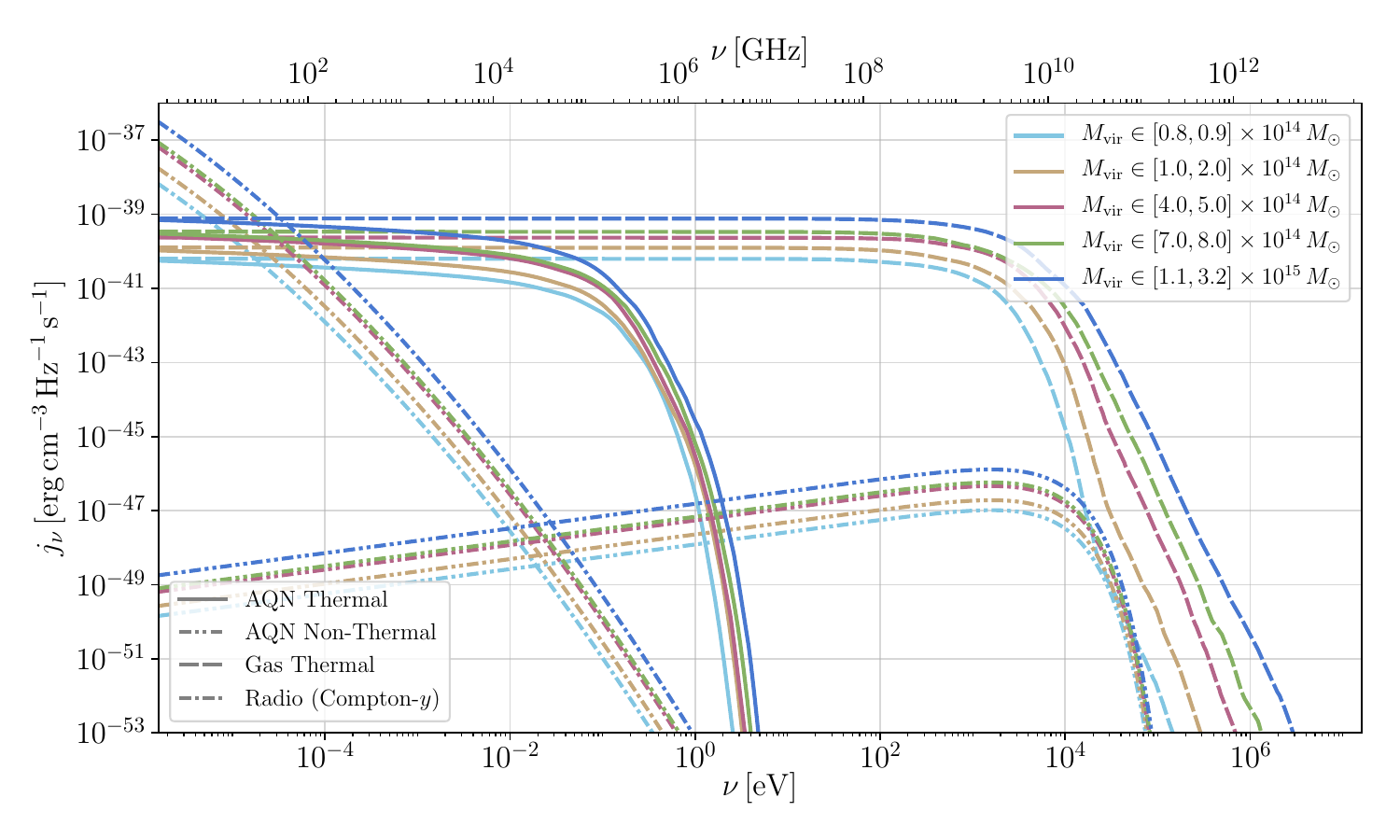}
    \caption{Median cluster spectra with lines for each mass bin out of \samplea{} over a frequency range of $\nu\in[10^{-5},10^8]/(2\pi)\,\ev$. The solid, dash-dot-dotted, dashed, and dash-dotted lines represent \aqn{} thermal emission, non-thermal emission, ICM Bremsstrahlung, and synchrotron emission, respectively.}
    \label{fig:median_spec}
\end{figure*}

A spectral energy distribution can be obtained by using \autoref{eq:thermal_spec_nat}, \autoref{eq:majidi_nt_taqn}, and \autoref{eq:xray_emissivity} for thermal and non-thermal \aqn{} emission, and thermal Bremsstrahlung emission from the ICM, respectively. For synchrotron emission, we followed the steps described in \autoref{sec:crs}. It shall be noted that we have to apply \autoref{eq:aqn_emissivity} to \autoref{eq:thermal_spec_nat} and \autoref{eq:majidi_nt_taqn}, to obtain an emissivity in units of $\emisunits$. The emissivity is then calculated for each particle in each frequency in a range of $\nu\in[10^{-5},10^8]/(2\pi)\,\ev$ with a spectral resolution of $10^3$ bins for \aqn{} and gas. For the CR synchrotron emission, frequency ranges of $\nu_\mathrm{synch}\in[10^{-5},10^{-2}]/(2\pi)\,\ev$ were chosen in a spectral resolution of $100$ bins. \autoref{fig:median_spec} displays the spectral energy distribution of the five mass bins from \samplea: Here, the median for each frequency in each mass bin was taken out of the 30 respective galaxy clusters. \aqn{} thermal emission, non-thermal emission, ICM Bremsstrahlung, and CR synchrotron emission are represented by solid, dash-dot-dotted, dashed, and dash-dotted lines, respectively.

It is noticeable that higher cluster masses result in an increment of the general spectral energy. In this figure, the reader's attention shall be drawn towards the regions where both of the \aqn{} emission become comparable to the ICM Bremsstrahlung emission in the low and high energy regime, as it is possible to obtain small frequency windows where \aqn{} emission would overtake the ICM Bremsstrahlung. In the low energy regime of $\nu_\mathrm{low} \lesssim 10^{-3}\ev$, \aqn{} and ICM emission roughly increase equally for higher $\mvir$. This opens the possibility that thermal \aqn{} radiation has the potential to outshine ICM Bremsstrahlung independently of cluster masses. The fact that AQNs can deliver a Bremsstrahlung intensity similar to hot X-ray gas in the radio regime is particularly interesting since in contrast to $\tgas$, $\taqn$ is not as strictly related to $\ngas$, which highly depends on the equation of state and the equilibrium condition.

On the other hand, we would like to point out that the spectrum was extended towards minimum frequencies of $\nu_\mathrm{min} = 10^{-5}/(2\pi)\,\ev$ to provide a low-frequency coverage by the synchrotron emission. For \aqns{}, however, it can become challenging to provide a physically meaningful interpretation for frequencies lower than $\nu_\mathrm{min}$, because of the finite size effect (see \cite{Forbes2008b}). Here, the plane wave approximation breaks down, as the positron wavelength becomes comparable to the size of the nugget.

It becomes a bit more tricky when comparing non-thermal \aqn{} emission to the high energy Bremsstrahlung regime in frequencies of $\nu_\mathrm{high}\gtrsim 10\,\kev$. Non-thermal \aqn{} emission will increment not as strongly as X-ray emission of the hot cluster gas. Since $\xrayemisint\sim \rhogas^2\tgas^{1/2}$ (see \autoref{eq:xray_emissivity}) and $\tgas$ increases with $\mvir$ (see \autoref{fig:histograms}), $\xrayemis$ will especially increase drastically for increasing $\mvir$ in the high energy regime, leading to the circumstance that non-thermal \aqn{} is expected to only be visible in sufficiently low cluster masses. 

The contribution in synchrotron emission in \autoref{fig:median_spec} is increasing for higher $\mvir$ because of the Compton-$y$ parameter. Synchrotron emission is likely to interfere with the thermal \aqn{} emission, especially in the regions where thermal \aqn{} emission would dominate the ICM Bremsstrahlung emission. It is therefore worth noting that one has to account for synchrotron emission and ICM Bremsstrahlung in the low energy regime when searching for an optimal frequency window for thermal \aqn{} emission.

In the following paragraphs, we will focus on the promising frequency bands where \aqn{} emission is expected to dominate. First, to measure how strongly \aqn{} emission outshines the background emission a metric has to be defined that includes the frequency and the magnitude of the increment in emission. Depending on the mass and other physical properties of the galaxy cluster, thermal and non-thermal \aqn{} emission will distinctively be higher or lower in individual frequencies for different galaxy clusters. Therefore, a transition frequency $\nutrans$ has to be defined, where the absolute summed emissivity of all \aqn{} tracer particles per cluster per frequency ($\aqnemis$) transitions from higher to lower values compared to the thermal gas emission. From a minimum frequency $\numin$ to the transition frequency all background emission sources (including synchrotron emission) will be integrated. In the simulation, the numerical integration is conducted by applying the trapezoidal rule which is implemented in the \package{NumericalIntegration.jl} package.\footnote{\url{https://github.com/dextorious/NumericalIntegration.jl}}

\subsubsection{Low Frequency Regime}

For the low energy regime, we define $\numin$ to be the lowest possible frequency that was used to calculate the spectra, \ie{} $\numin = 10^{-5}/(2\pi)\,\ev$ because of the finite size effect of \aqns. After integration from $\numin$ to $\nutrans$, all background sources will be summed to represent a total galaxy cluster emission. The integrated spectral ratio of \aqns{} $\specratio$ significantly contributing to the background emission can be calculated by:

\begin{equation}\label{eq:all_specdiff_low_energy}
    j_\mathrm{tot}=\underbrace{\int_{\nu_\mathrm{min}}^{\nu_T}j_{\nu,\mathrm{AQN}}\,\diff\nu}_{=:j_\mathrm{AQN}} + \int_{\nu_\mathrm{min}}^{\nu_T}j_{\nu,\mathrm{gas}}\,\diff\nu + \int_{\nu_\mathrm{min}}^{\nu_T}j_{\nu, \mathrm{synch}}\,\diff\nu,
\end{equation}

\noindent
and

\begin{equation}\label{eq:all_specratio}
    \specratio=\frac{j_\mathrm{AQN}}{j_\mathrm{tot}}.
\end{equation}

In the following analysis, the $\specratio$ was determined in both the low and high energy regimes for \samplea{} and \sampleb{}. After applying the condition that thermal \aqn{} emission must exceed thermal gas emission at frequencies below $\nutrans$, approximately $40.99\%$ of all galaxy clusters combined from \samplea{} and \sampleb{} exhibit stronger \aqn{} emission compared to the thermal background gas emission from $\nu_\mathrm{min}$ to $\nutrans$.

\begin{figure*}
    \centering
    \includegraphics[width=\linewidth]{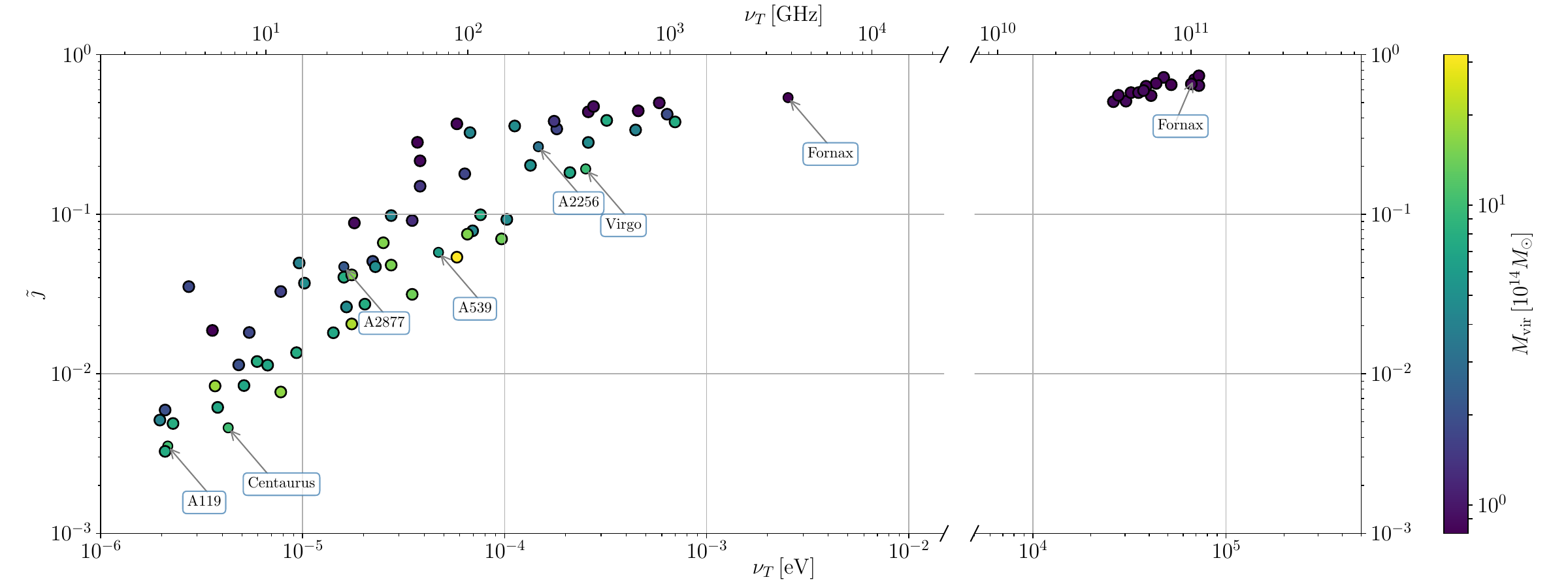}
   \caption{
    Transition frequency-integrated ratio-relation, color-coded by $\mvir$ with frequency ranges of positive \aqn{} contributions. The low-energy frequency range shows in which the thermal \aqn{} emission dominates. Low cluster masses tend to reside on a tangent, slightly offset towards higher values. The high-energy frequency range shows in which the non-thermal \aqn{} emission dominates. Only \aqn{} emission, originating from the lowest $\mvir$ shows an excess over the thermal gas emission in the high-energy regime. We removed one cluster from the plot where for numerical reasons we could not compute $\nu_T$ and the associated $\specratio$ reliably.}
    \label{fig:intspecratio}
\end{figure*}

Using this subset of galaxy clusters and applying \autoref{eq:all_specratio}, produces \autoref{fig:intspecratio}, which displays the integrated spectral ratio as a function of the transition frequency for both frequency windows where \aqn{} emission exhibits an excess. On the left side of the figure the thermal regime of the \aqns{} dominates, whereas on the right side at high-energy frequencies, the non-thermal \aqn{} regime dominates. An important outcome is that galaxy clusters with relatively low $\mvir$ are still capable of reaching high $\specratio$ and follow relatively well-constrained tangents indicated by the colormap for different cluster masses. 

It is important to find out how cross-identified galaxy clusters populate in this relation to identify promising real-world cluster candidates. It is therefore shown in \autoref{fig:intspecratio} how \sampleb{} is distributed in the $\nutrans$-$\specratio$ diagram. It can be seen in the low-energy regime that Fornax and Virgo appear to be the best candidates for the strongest \aqn{} emission offset for the background gas emission while extending to the highest $\nutrans$. Even though it is the most massive galaxy cluster out of \sampleb{}, the Coma cluster due to its very strong, observed synchrotron emission, does not satisfy the condition to exhibit a stronger \aqn{} emission over thermal gas emission in the energy window of $\nu_\mathrm{min}$ to $\nutrans$ and is therefore not displayed in \autoref{fig:intspecratio}.

\begin{table*}
	\centering
    \caption{Table of the cross-identified galaxy clusters with physical properties from the simulation and their ability to show \aqn{} signatures with and without synchrotron radiation in the background ($\Delta j_\text{no synch}$, $\Delta j_\text{with synch}$) at a transition frequency $\nutrans$ in the low-energy regime.}
	\begin{tabular}{l c c c c}
        \toprule
        Cluster Name & $\mvir$ & $\nutrans$ & $\tilde{\jmath}$ (no synchrotron) & $\tilde{\jmath}$ (with synchrotron) \\
        & $[10^{14}\,M_\odot]$ & $[\GHz]$ & $[\%]$ & $[\%]$ \\
        \midrule
        Coma & $18.82$ & \xmark & \xmark & \xmark \\
        Virgo & $9.91$ & $382.45$ & $53.2$ & $19.22$ \\
        Perseus & $10.75$ & \xmark & \xmark & \xmark \\
        Centaurus & $10.18$ & $6.5$ & $50.97$ & $0.46$ \\
        A119 & $10.21$ & $3.26$ & $50.33$ & $0.35$ \\
        A539 & $6.34$ & $71.42$ & $52.43$ & $5.76$ \\
        A1185 & $4.42$ & \xmark & \xmark & \xmark \\
        A2256 & $3.27$ & $223.02$ & $53.17$ & $26.49$ \\
        A2877 & $2.22$ & $24.29$ & $51.87$ & $4.69$ \\
        Norma & $0.79$ & \xmark & \xmark & \xmark \\
        Fornax & $0.61$ & $3842.19$ & $54.33$ & $53.76$ \\
        \bottomrule
	\end{tabular}
    \label{tab:digtwin_params}
\end{table*}

\subsubsection{High Frequency Regime}

Since non-thermal \aqn{} emission is expected to show two intersections in the high energy regime, one has to define $\numin$ to be the first transition frequency where $\ntemis > \backgroundemis$ and $\nutrans$ is again where $\ntemis < \backgroundemis$ with the condition that $\numin < \nutrans$. Non-thermal \aqn{} emission causing an excess over the background emission in the high energy regime is mostly dictated by the individual X-ray brightness of a galaxy cluster (see \autoref{fig:median_spec}).

It can be seen on the right-hand side of the broken axis in \autoref{fig:intspecratio} that only a small fraction of galaxy clusters could potentially dominate background gas emission with \aqn{} emission. When applying the conditions to determine an excess in the high energy regime by \aqns{}, $9.32\%$, \ie{} a total number of 15 out of both samples was identified. \autoref{fig:high_energy_spec} depicts the frequency window in which thermal gas emission is expected to intersect with the non-thermal \aqn{} emission. In this image, dashed lines represent the Bremsstrahlung emission from the ICM and dash-dot-dotted lines are the non-thermal \aqn{} emission. If there is an excess occurring, the signature is only very subtle and many low-mass galaxy clusters are still too bright in X-rays to permit a window for non-thermal \aqn{} emission to dominate. When comparing both regimes in \autoref{fig:intspecratio}, it can be seen that in the high-energy regime, an \aqn{} excess can only be visible for low cluster masses, which then results in a dimmer gas emission. Galaxy clusters in a higher dynamical state of dark matter particles to the surrounding gas particles, could additionally positively contribute to a non-thermal \aqn-excess, since $\ntemis \sim \naqn\,\ngas\,\dvel$.

It shall be noted that the non-thermal emission model is not as robust as the thermal emission model of the \aqns{}. In future studies, the exponent of $1/3$ in \autoref{eq:majidi_nt_reff} and \autoref{eq:majidi_nt_taqn} for the frequency, will most likely be replaced by a free parameter $\beta$ allowing to be adaptive. The exponent of $1/3$ is an approximate solution to solve the synchrotron function. It is crucial to note that particular solutions for the synchrotron function to model the non-thermal spectrum have to be motivated in greater detail and new precise models should aim for a more detailed physical implementation. In addition, the critical frequency as proposed in \autoref{subsec:aqn_rad} is set to $30\,\kev$ by convention and should be understood in an \enquote{order of magnitude estimate} \citep{Forbes2008a}. To provide improved estimates, a refined theory of the non-thermal physics in the structure of \aqns{} is desired. The current model utilizes an approximate treatment in the mean-field of the Maxwell equations. It is therefore not yet possible to discard the possibility of detecting \aqn{} signatures in the high energy regime immediately and a final conclusion shall be made once a better non-thermal \aqn{} model is developed.

\subsection{Cluster Maps} \label{sec:cluster_maps}

\begin{figure*}
    \centering
    \includegraphics[width=\linewidth]{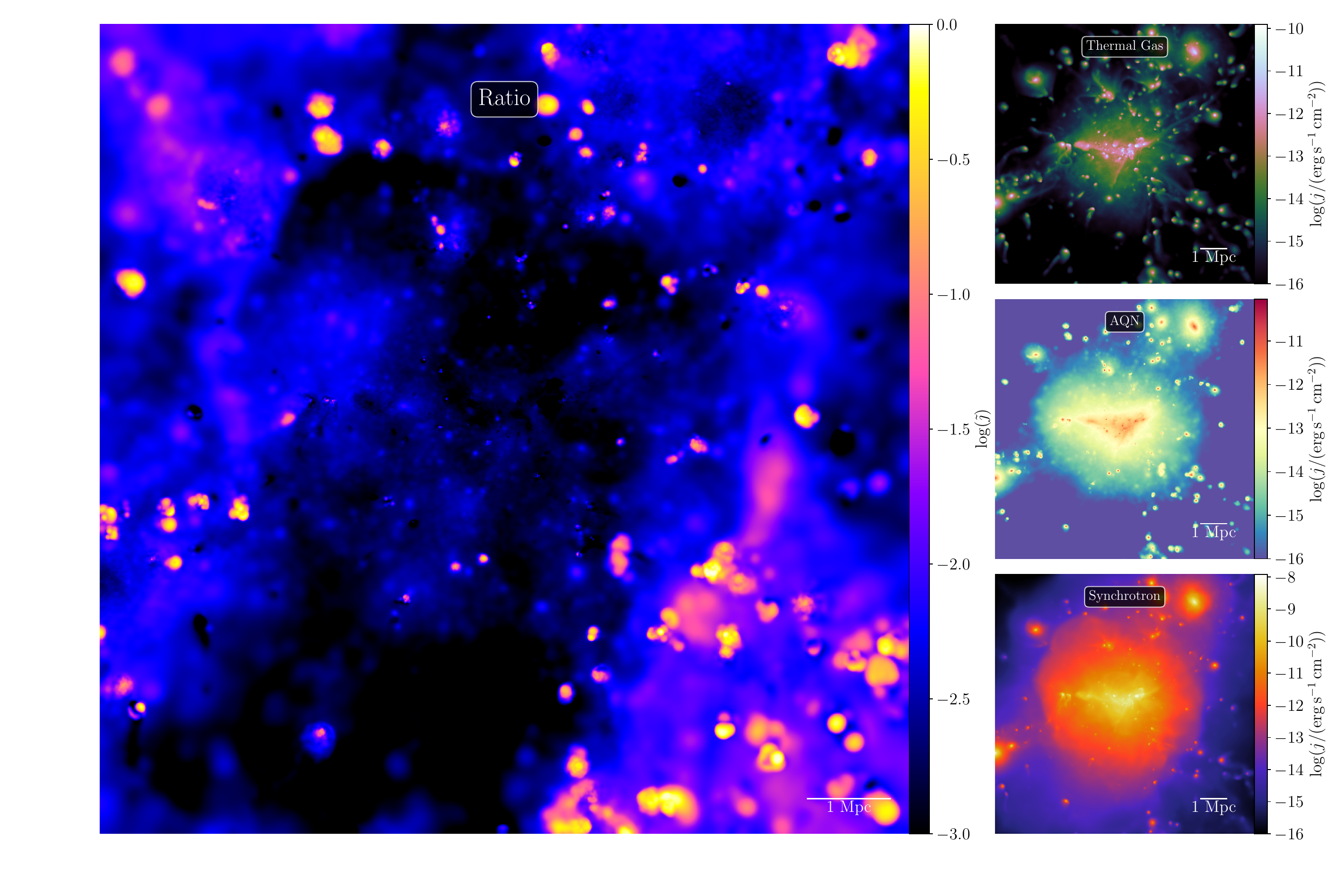}
    \caption{The left panel shows the ratio-map expressed in $\specratio$, and the right column shows from top to bottom, the thermal gas, AQN and synchrotron emission of Coma integrated over $1.40\,\GHz$ to $4.85\,\GHz$, projected $3\,\rvir$ along the line of sight. In this frequency window, the synchrotron emission of Coma is well constrained through radio emission \citep{Thierbach2003coma}. Due to the strong radio emission of Coma, here the contribution to the diffuse emission by AQNs is in the sub-percent level.}
    \label{fig:coma_maps}
\end{figure*}

It is important to note that one loses information when only considering spectral features. Since \aqns{} spatially follows the halo distribution, positions of dark matter with respect to gas might slightly differ from each other. Extremely low emissions do not contribute to the spectrum, because $j_\nu$ was summed over all particles in the corresponding frequency bin, so spatial regions of low emission are not resolved in a spectrum. It is especially important for the gas emission to identify regions of low emissivity where \aqn{} emission would be proficient to outshine thermal gas emission. It can even be possible to find an \aqn{} excess in small regions in spatially resolved maps where one did not expect an excess only by considering the spectrum. On the other hand, it is not possible to look for excess features in all 161 galaxy clusters by finding their individual optimal region where \aqn{} features dominate over the background emission when scanning through a spectral resolution of 1000 frequency bins. We, therefore, stick to the methodology presented in \autoref{eq:all_specdiff_low_energy} to analyze cluster regions, where \aqn{} radiation certainly dominates in the spectrum by inferring the emission offset and only consider promising candidates of cross-identified galaxy clusters. Note that for simplicity the maps for all clusters are done in supergalactic $x$/$y$ coordinates and not rotated to match a real observer in the Local Universe.

\begin{figure*}
    \centering
    \includegraphics[width=\linewidth]{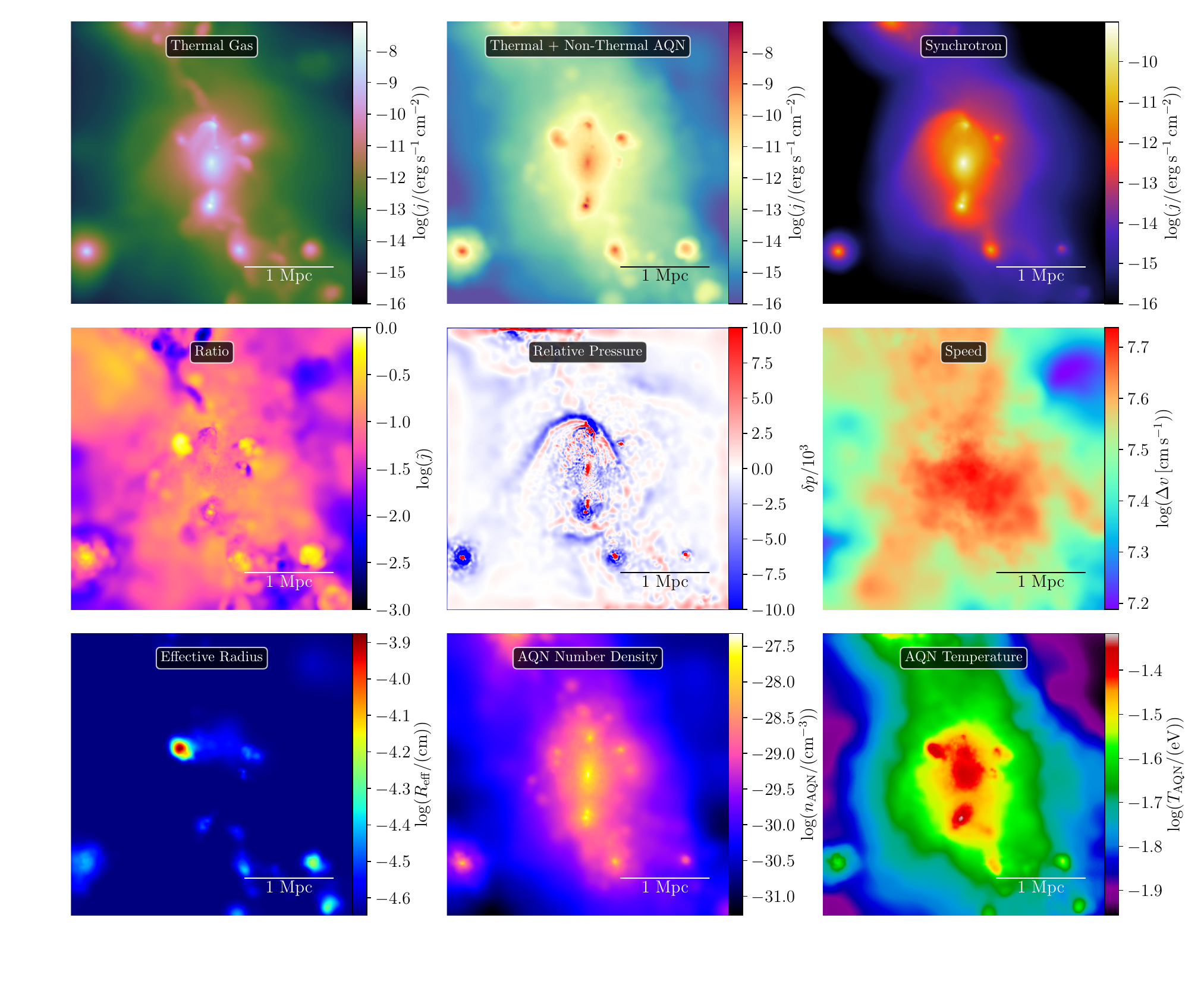}
    \caption{Cross-identified Fornax cluster represented in different relevant properties.}
    \label{fig:fornax_maps}
\end{figure*}

\begin{figure*}
    \centering
    \includegraphics[width=\linewidth]{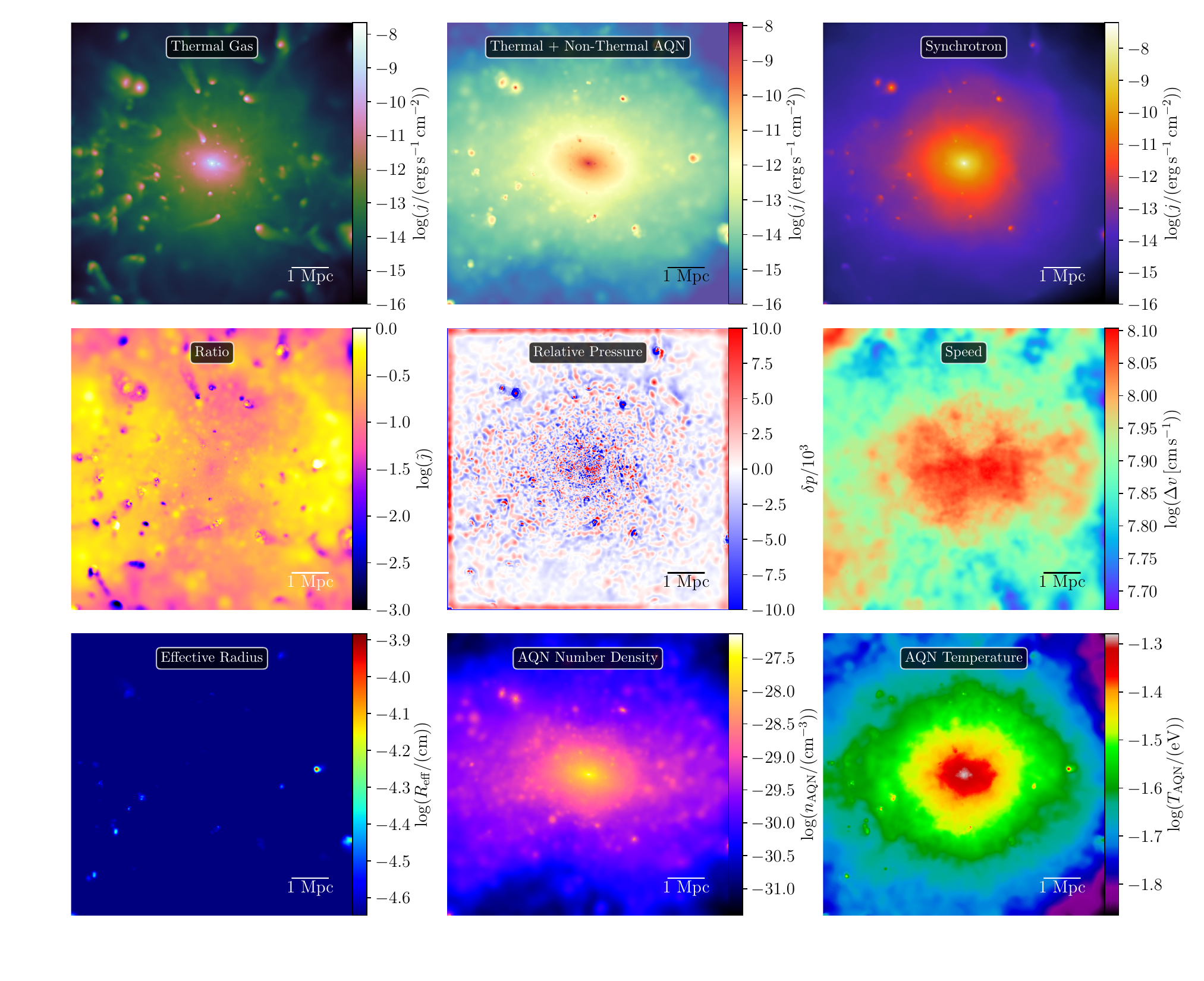}
    \caption{Cross-identified Virgo cluster represented in different relevant properties.}
    \label{fig:virgo_maps}
\end{figure*}

Given this reasoning, in this section, SPH maps of gas and \aqns{} are represented using their most contributing physical properties and fluxes integrated over an adaptive best possible frequency range. \autoref{fig:fornax_maps} and \autoref{fig:virgo_maps} display the Fornax and Virgo clusters in a selection of relevant properties. 

Relative pressure maps were obtained by generating SPH maps by utilizing the ideal gas relation $\pgas \sim \ngas\tgas$. Hence, the product of $\ngas$ and $\tgas$ will represent the scaling of the pressure in the gas particles and SPH maps were constructed this way. With the obtained pressure map, a Gaussian filter was applied to the projected bitmap of the galaxy cluster with $\sigma=5$ in the Gaussian kernel. The relative pressure change is therefore obtained via:

\begin{equation}
    \delta p = \frac{p_\mathrm{original}-p_\mathrm{gaussian}}{p_\mathrm{gaussian}}
\end{equation}

In Fornax (see \autoref{fig:fornax_maps}), less filamentary structures with strong emission are visible in thermal \aqn{} emission compared to thermal gas emission, because dark matter is not affected by environmental friction effects that would perturb the particle distribution. In the map depicting $\specratio$, the strongest contributions originate from a larger infalling substructure, which is visible in the effective radius map as well. The $\specratio$-map specifically shows the presence of the glowing axion quark nugget component, which seems to show strong emissions in regions of the ICM.

Prominent regions of $\reff > \raqn$ are only possible if the gas environment provides the necessary parameter combination. Compared to other cross-identified galaxy clusters, it is rather rare to see large regions with high $\reff$ values. The effective radius increments with increasing $\ngas$ and $\dvel$ and decrements for increasing $\tgas$. $\tgas$ scales with the largest exponent and therefore shows the strongest influence on $\reff$ with especially high gas temperatures in the central regions. Fornax is a relatively small cluster with a comparable low thermal gas emission coming from a low $\tgas$. $\reff$ can therefore reach values of $\reff > \raqn$ in central regions, too. However, when comparing $\reff$ maps in Fornax to Virgo and Coma, one can see that $\reff > \raqn$ values are more likely to be found in the peripheries of galaxy clusters.

An intriguing physical property is that regions of $\reff > 10^{-4}\,\cm$ are typically not abundant in relative velocity maps, and it is especially in these regions where $\dvel$ seem to be lower compared to their nearest surroundings. Following $\reff\sim \ngas^{5/7}\dvel^{5/7}\tgas^{-17/7}$ (cf. \autoref{eq:reff_full}), gas overdensities in conjunction with low $\tgas$ yield higher $\reff$. In the case of Fornax, the prominent $\reff> \raqn$-region is too close to the cluster center and will therefore be hotter than in the outskirts. In this case, $\ngas$ would be the only free parameter that could directly influence $\reff$.

Regions of high $\taqn$ do not necessarily follow the $\naqn$ distribution, since $\taqn$ is also influenced by the surrounding gas. Shocked regions are not significantly embedded in the \aqn{} emission map and vice versa. A further interesting feature is that the \aqn{} emission does not seem to show a significantly stronger abundance at extended regions. This is not trivial since dark matter particles do not suffer from friction and do not interact other than with gravity in the simulation. Therefore, it would have been sensible to expect additional emission contributions in the periphery of Fornax that are not strongly present in the gas emission maps.

The Virgo cluster (see \autoref{fig:virgo_maps}) differs strongly from Fornax as it is more massive, hotter and in a more relaxed state. No signs of a current major merger are visible in the relative pressure maps (other than visible in Fornax with its prominent northern bow shock). And yet, both of these clusters independently show the strongest \aqn{} excess. It is not obvious that shock features are no tracers for \aqn{} emission as shocked regions are always accompanied by significant offsets in thermodynamic properties (such for instance density or temperature). It was therefore expected that at least shocked regions will play a diminishing or reinforcing role on the \aqn{} emission --  in neither of the galaxy clusters this can be directly confirmed, and shock maps of other cross-identified galaxy clusters do not show signs of shock signatures in the \aqn{} maps either.

In the $\specratio$-map of the Virgo cluster, one can see that substructures show a mix of stronger and weaker contributions from the \aqn{} emission to the ambient background emission. High $\dvel$ in the outskirts of Virgo have an impact on larger values of $\specratio$ in the peripheries. This direct influence, however, is not obvious in the case of Fornax. In the Virgo cluster, the strongest contribution of $\specratio$ seems to originate from the regions of the ICM instead of the infalling substructures. This suggests looking for \aqn{} signatures where no galaxies are located when searching for signatures outside the center of the galaxy cluster.

In both clusters, one can see that high $\dvel$ can also be reached in the clusters' peripheries. The distribution of strong $\dvel$ appears to be anisotropic and \aqn{} emission does not necessarily follow the distribution of high relative velocities on large scales.

To learn how \aqn{} signatures evolve in bands of higher energies, individual energy ranges of four different instruments were taken. \aqn{} and gas emission were integrated within the respective energy ranges and \aqn{} emission was superimposed by its thermal and non-thermal components. For the sake of versatile energy coverage, the instruments WMAP \citep{Bennett2013}, Planck \citep{Tauber2010}, Euclid \citep{Laureijs2011} and XRISM \citep{Mori2022} were considered. \autoref{tab:instruments} shows the corresponding maximum energy range with upper and lower limits denoted by $\nu_\mathrm{min}$ and $\nu_\mathrm{max}$ for each instrument in units of $\GHz$.

\begin{table}
	\centering
	\begin{tabular}{c | c c}
        \toprule
        & $\nu_\mathrm{min}\,[\GHz]$ & $\nu_\mathrm{max}\,[\GHz]$ \\
        \midrule
        WMAP & $22.69$ & $93.43$   \\
        Planck & $30$ & $857$ \\
        Euclid & $1.50\times 10^5$ & $5.45\times 10^5$ \\
        XRISM & $9.67\times 10^7$ & $3.14\times 10^9$ \\
        \bottomrule
	\end{tabular}
    \caption{Band-passes for WMAP \citep{Bennett2013}, Planck \citep{Tauber2010}, Euclid \citep{Laureijs2011} and XRISM \citep{Mori2022} in units of $\GHz$.}
    \label{tab:instruments}
\end{table}

\begin{figure*}
    \centering
    \includegraphics[width=\linewidth]{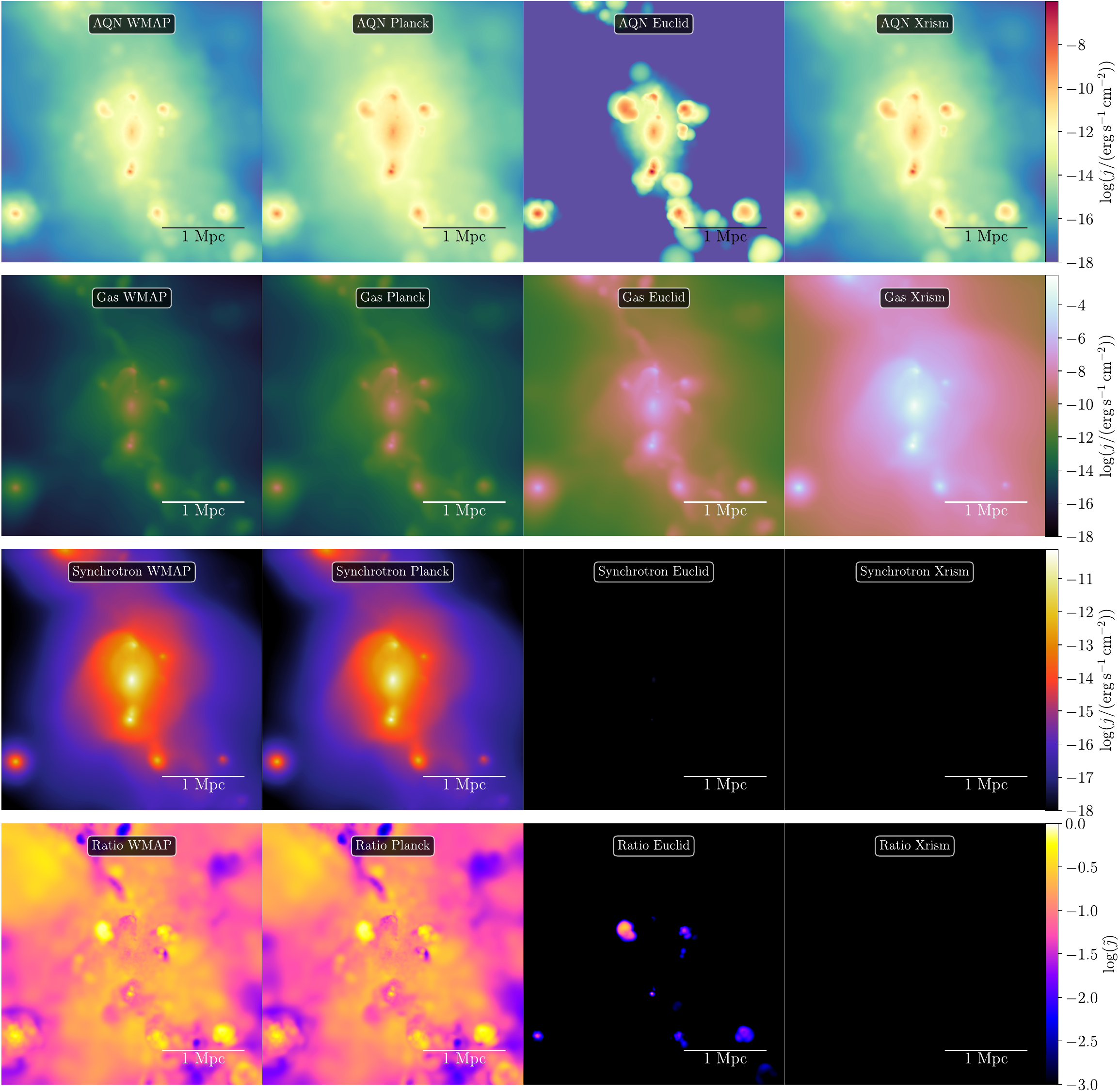}
    \caption{Fornax cluster \aqn, gas and synchrotron emission generated from WMAP, Planck, Euclid and XRISM energy ranges. Synchrotron emission is predominantly abundant in WMAP and Planck bands. Thermal gas emission shows the most significant contribution to the background emission in Euclid and XRISM energy bands.}
    \label{fig:fornax_bands}
\end{figure*}

\begin{figure*}
    \centering
    \includegraphics[width=\linewidth]{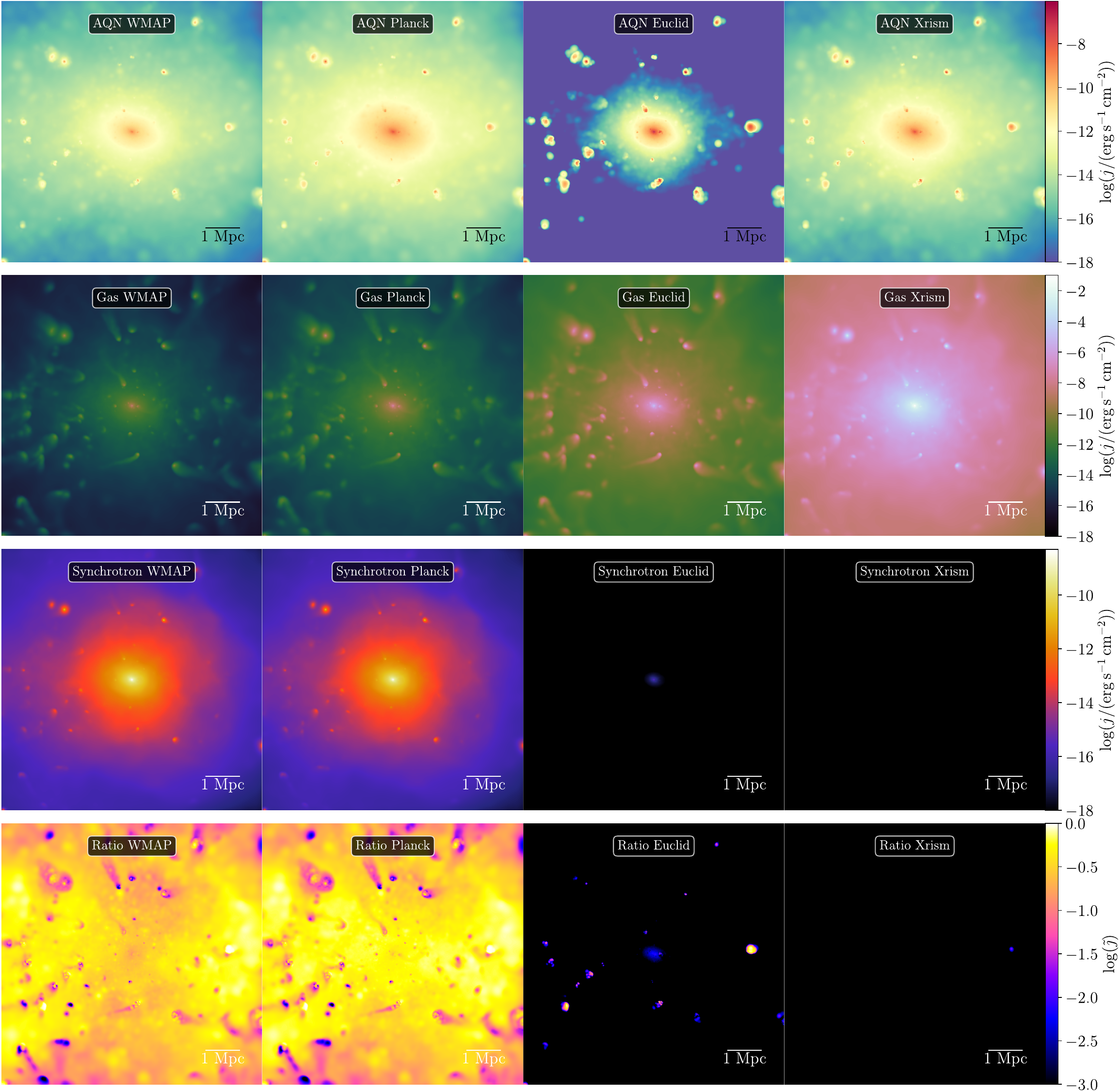}
    \caption{Virgo cluster \aqn, gas and synchrotron emission generated from WMAP, Planck, Euclid and XRISM energy ranges. Synchrotron emission is predominantly abundant in WMAP and Planck bands. Thermal gas emission shows the most significant contribution to the background emission in Euclid and XRISM energy bands.}
    \label{fig:virgo_bands}
\end{figure*}

In the following the two most promising galaxy clusters, determined by the top figure in \autoref{fig:intspecratio} were selected for a multiband analysis. \autoref{fig:fornax_bands} and \autoref{fig:virgo_bands} show the gas and \aqn{} emission with the corresponding ratio image in the selected bands. To compare how the brightness of the emission evolves, all frames in their corresponding emission feature share the same colorbar limits.

While the gas emission shows a continuing brightening for increasing band-energies, the \aqn{} signatures show a significant decrease in brightness in the Euclid band. This decrement originates from the spectral feature that thermal \aqn{} emission transitions into the non-thermal regime. Euclid operates in the energy range, where the thermal emission experiences its cutoff (\cf{} \autoref{fig:median_spec}). Even though \aqn{} signatures show a re-brightening once the non-thermal regime takes over, \aqn{} emission cannot compete with the thermal gas emission in the X-ray regime. It is visible in the ratio-images that \aqn{} emission would only be able to dominate in the low energy regime. A re-brightening at higher energy bands is only visible in infalling substructures, however, the ambient glow is not abundant anymore.

It can be concluded that Euclid and XRISM are the least promising instruments for a potential \aqn{} signature detection as the thermal gas emission dominates over the \aqn{} emission in this energy range. WMAP and Planck, exhibit \aqn{} signatures in the ratio-images, even when combining thermal gas emission with synchrotron emission.

\subsection{Observational Feasibility of Promising Cluster Candidates}

It became clear in \autoref{sec:cluster_maps} that spatial features of strong \aqn{} signatures are hard to identify and differ from cluster to cluster and their dynamical states. Even though \aqn{} emission excess is abundant in galaxy clusters, it becomes challenging to pinpoint regions of clusters that are expected to be the most promising.

In observations, galaxy cluster emissions are typically analyzed by radial profiles in a given frequency band. This approach enables the identification of important spatial features that are dependent on distance and the corresponding frequency band. In this section, radial profiles will be shown for the cross-identified galaxy clusters that show an excess signature of thermal \aqn{} emission integrated over their individual frequency bands from $\numin$ to $\nutrans$. Radial profiles in 50 bins were taken for each emission property and a final radial profile was calculated after summing the \aqn, gas and synchrotron maps.

\autoref{fig:emradprof} shows how the $\specratio(r)$ evolves for the cross-identified galaxy clusters over a radius of $r\in [10^{-2},1.5]\,\rvir$. An interesting feature in the radial profiles is that galaxy clusters exhibiting high $\specratio$ in \autoref{fig:intspecratio} also show high ratios in their central regions. In comparison to the remaining cross-identified galaxy clusters, Fornax and Virgo stand out in \autoref{fig:emradprof}.

In all radial profiles, the most significant contribution of thermal \aqn{} emission comes from the peripheral regions, where thermal gas and synchrotron emissions have sufficiently decreased.

\section{Discussion}\label{sec:discussion}

Numerical analysis of \aqns{} in galaxy clusters in a cosmological simulation has shown that direct signatures of this new dark matter model are possible, but can be challenging to pinpoint. Spatial distributions of significant \aqn{} emission contributions depend on factors, like the dynamical cluster state and thermal \aqn{} emission is spectrally superimposed with non-thermal synchrotron and thermal gas emission.

Given that the spectral AQN signature is not exactly bremsstrahlung and differs from synchrotron, component separation -- where templates are used and spectral signatures help disentangle sources of emission -- could be a reasonable strategy to look for weak \aqn{} contributions. Multifrequency data would especially improve this approach (see \cite{Majidi2023} for a discussion).

In \autoref{sec:interactions}, tracers of \aqn{} signatures and further unexplained observational phenomena were pointed out and analyzed by various studies. The study by \cite{Lawson2019a} proposed that thermal \aqn{} emission might be responsible for a stronger absorption feature in the $21\,\cm$ line at a redshift of $z=17$ during the early universe. Clusters of galaxies have not formed at this time yet and early protoclusters -- observationally confirmed to be progenitors of present-day galaxy clusters -- are detected at $z\sim 6-7$ \citep{Harikane2019}. Cosmological, hydrodynamical simulations suggest that a hot ICM is already present at $z=4.3$, at least within the core of such proto cluster \citep{remus2023proto}. We therefore cannot assume the same conditions for the \aqn{} increment which was obtained in the framework of this paper. However, it is worth mentioning that protoclusters at $z\sim 7$ could be capable of influencing the radio background if the \aqn{} signal would be similarly significant as in galaxy clusters at present time. An increment in \aqn{} emission was concluded in \cite{Majidi2023} after constructing a light cone to a redshift of $z=5.4$ using the \emph{Magneticum}\footnote{\url{http://www.magneticum.org/}} simulation \citep{Dolag2016}. The sum of the \aqn{} emission increment in the radial profiles from the cross-identified galaxy clusters of our work at $z=0$ can be attributed to $\sum \specratio(r) \approx 4.80\,\%$ thermal \aqn{} emission.

It is however important to mention that early clusters possess different physical properties \citep{Chiang2013, Overzier2016}, which would likely yield different \aqn{} signatures. Results from this study only provided information of \aqn{} signatures from cluster properties at $z=0$. It would be sensible to implement \aqn{} features already in snapshots at higher redshift and develop a dynamical \aqn-treatment, to grasp a full evolution of \aqn{} signatures from early to present-time galaxy clusters that could contribute to the background radio emission.

The study by \cite{Zhitnitsky2022a} analyzed the diffuse galactic UV radiation and suggested that hot \aqns{} with $\taqn\sim 5\,\ev$ could reproduce such observations -- however since our study focuses on scales of galaxy clusters and not galactic scales, we cannot verify these predictions. The same argumentation is applied to the proposal to the contribution of Chandra's diffuse $8\,\kev$ emission by \aqns{} \citep{Forbes2008a}. However, it is worth commenting that in a galaxy cluster, it is rather unlikely to find \aqns{} with temperatures of $5\,\ev$ because of the suppression of capturing protons in the hot ICM. Galactic environments differ strongly from galaxy cluster environments with different $\tgas$, $\ngas$, $n_\mathrm{DM}$ and $\dvel$ as the interstellar medium (ISM) is cooler and denser than the ICM. In addition, galaxies exhibit a significantly lower gas emission and especially non-thermal \aqn{} emission in the high energy regime might indeed be capable of dominating highly energetic galactic background emission.

It is not possible to comment predictions on the $511\,\kev$, as predicted by \cite{Oaknin2005a, Zhitnitsky2007, Forbes2010a, Flambaum2021} as we did not consider $e^+e^-$ annihilation lines in our approach.

In \autoref{sec:cluster_maps}, \aqn{} emission was studied for different instruments in their corresponding bands. While this paper focused on \aqn{} properties in galaxy cluster environments rather than properties within the Milky Way, a direct comparison to \cite{Forbes2008b} cannot be conducted. Nevertheless, it is striking to observe the abundance of \aqn{} signatures in the WMAP maps depicted in \autoref{fig:fornax_bands} and \autoref{fig:virgo_bands}. \aqn{} signatures in the WMAP band-pass can be identified even after including synchrotron background emission. While it is important to interpret the results with caution given the significant differences between cluster and galactic properties, the possibility of an excess in microwaves in the galactic core due to \aqn{} emission may not be entirely unfounded.

\section{Conclusions}\label{sec:conclusion}

We studied the glow of axion quark nugget dark matter as signatures in a large sample of 161 simulated galaxy clusters galaxy with a constrained cosmological simulation of the local universe (SLOW). We provide lower-limit predictions on electromagnetic counterparts of AQNs in the environment of galaxy clusters by inferring their thermal and non-thermal emission spectrum originating from axion quark nugget-cluster gas interactions. Dividing our cluster sample into a sub-sample ordered in five mass bins ranging from $0.8$ to $31.7 \times 10^{14} \,\msol$ allows us to specify the contribution of the \aqn{} signal to the emission of the ICM as a function of cluster mass. Having 11 cross-identified galaxy clusters within this sample allows us in addition to propose promising galaxy cluster candidates, where the AQN signal might show the strongest contribution to the overall emission.

\subsection{Key Results}
Throughout this analysis, we fixed the parameters of the underlying AQNs model to have $\maqn = 16.7\,\g$. We then computed effective radius ($\reff$) and temperatures ($\taqn$) of the AQNs based on the local gas and dark matter properties within the simulations, which allowed us to compute the expected electromagnetic signal of the AQNs as well as the thermal and non-thermal emission of the ICM, where for the later we assumed the same ratio of between cosmic ray electrons and thermal energy as observed in the Coma galaxy cluster. Our key findings can be summarized as follows:

\begin{enumerate}
    \item Even though more massive galaxy clusters typically exhibit higher ICM temperatures, the largest population of high $\reff$ can be observed for the most massive mass bin in \samplea{} as $\dvel$ scales with cluster mass as well. This is an interesting feature since $\taqn$ directly scales with $\reff$ and the \aqn{} emission highly depends on $\taqn$. Consequently, strong \aqn{} emission can be expected for massive galaxy clusters. The increased \aqn{} emission, however, is always accompanied by high cluster gas emission, as thermal gas emission of the ICM scales with cluster gas temperatures as well. To some degree, relative emission domination is therefore equilibrated for higher cluster masses.
    \item A strong \aqn{} emission contribution scales with the mass of a galaxy cluster in the low energy regime. Galaxy clusters with different $\mvir$ can be populated at a fixed $\nutrans$ and differences in the absolute value of $\specratio$ are attributed to a combination of $\tgas$, $\ngas$ and $\dvel$.
    \item In the low-energy frequency range, approximately $40.99\%$ of galaxy clusters show a stronger emission from $\nu_\mathrm{min}$ to $\nutrans$ in the \aqn{} emission over thermal gas emission. After including synchrotron emission, values of the integrated spectral ratio $\specratio$ are distributed in the range of $0.33\%$ to $53.76\%$. However, as not all galaxy clusters feature giant radio halos, this opens a window of opportunity when the synchrotron emission of individual galaxy clusters is sufficiently low.
    \item Only low-mass galaxy clusters showed an excess of non-thermal \aqn{} emission attributing approximately $9.32\%$ to the 161 galaxy clusters out of both samples with values of $\specratio$ being distributed in between $50.71\%$ and $73.65\%$.
    \item A small amount of galaxy clusters exhibit large values of $\nutrans$. This enables to conduct follow-up observations in a wider range of possible surveys. Especially Fornax and Virgo host environments which are responsible for extremely high transition frequencies of $\nutrans = 3842.19\,\GHz$ and $382.45\,\GHz$ respectively. To make use of this feature that is naturally embedded in the thermal emission property of \aqns, improvements on pinpointing regions in galaxy clusters that can only be attributed to an \aqn{} emission excess are required.
    \item Shock features are no tracers for regions of strong \aqn{} emission.
    \item The spatial distribution of $\dvel$ in the SPH maps is not strongly correlated with \aqn{} emission signatures.
    \item Fornax and Virgo show in the WMAP and Planck $\specratio$-maps significant \aqn{} signatures. Due to their different dynamical states, regions of strong \aqn{} emission vary in their spatial distribution. Fornax, as the dynamical cluster has strong \aqn{} emission contributions in the infalling substructures, whereas Virgo predominantly shows high $\specratio$-values originating from locations of the ICM, and distributed in both central and peripheral regions of the cluster.
    \item Even though a general increase of emission in the radial profiles can be attributed to \aqn{} signatures, it is hard to identify specific regions that only trace \aqn{} signatures, since radial \aqn{} emission profiles share a similar morphology as gas emission profiles. We assumed the model to be true and showed that \aqn{} emission affects the spectrum and is visible in the emission maps. The \aqn-background emission offset, however, is so small that it is not possible to verify or discard this dark matter model observationally, even though it might be one of the most promising ones.
\end{enumerate}

\clearpage

\subsection{Future Work}

First of all, our study was constrained to regions enclosing a radius of $1.5\,\rvir$. This provides first valuable insights on expectable \aqn{} signatures from the innermost regions. However, it is expected that especially in the peripheries of galaxy clusters -- more specifically in the warm-hot intergalactic medium -- the environmental gas temperature of $\tgas\in[10^5,10^7]\,\kelvin$ \citep{Dave2001} may permit a better detection of \aqn{} signatures. Lower gas temperatures and number densities would yield a lower thermal gas emission. On the other hand, a lower number density of gas and \aqns{} would play a decrementing role in peripheral \aqn{} excess detections. Simultaneously, the background emission polluting the \aqn{} signature would decrease as well and both of the emissions will probably fall below the detection limit. However, it is still important to study the outskirts of galaxy clusters for tracers of \aqns{}, for example in simulations, to confirm or discard possible non-detection predictions.

Furthermore, we did not use a mass and size distribution for \aqns{} throughout this study and set every \aqn{} to a size of $\raqn= 2.25\times 10^{-5}\,\cm$ and a mass of $\maqn = 16.7\,\g$. Of course, this is a simplification that is not expected in the real world (see for example \cite{Ge2019}). It is reasonable to assume the utilized $\raqn$ and $\maqn$ as a rough reference value in an analysis conducted for a single snapshot at $z=0$ from the cosmological simulation. However, it would also be sensible to implement an on-the-fly model for \aqns{} directly evolving in a cosmological simulation. This would imply different halo distributions due to dynamical friction of differently massive \aqn{} particles, leading to a segregation of more massive \aqns{} in the central regions of galaxy clusters and probably influencing the radial dark matter density profiles, too. Consequently, not only density profiles would be altered by a constant influence of \aqns{} on the cluster environment throughout time. Annihilation and heating processes of \aqns{} would result in feedback mechanisms that act directly on the ICM. Therefore, temperature profiles in galaxy clusters would be influenced too, if the \aqn{} model in the simulation would be treated dynamically. An additional consequence would be that the heated ICM will become visible as tail-like traces behind substructures that fall towards the center of the galaxy cluster. Tails of galaxies are already detected in radio observations \citep{Vallee1988, Sun2005, Terni2017, Chen2020, deVos2021, Hu2021, Mueller2021, Pal2023} and are mostly attributed to AGN feedback mechanisms and ram-pressure stripping. Constant gas interactions with an infalling subhalo would cause similar effects that are recommended to be verified in a dynamical \aqn{} treatment in the simulation. A suitable \aqns{} mass distribution would therefore be reasonable if on-the-fly \aqn{} models are implemented in the simulations.

Throughout this paper, the similarity between \aqns{} and neutron stars was pointed out multiple times, as both objects are expected to host cores in a \cs{} state -- the same connection can be drawn to magnetars. It is assumed that strong magnetic fields of magnetized neutron stars can be established if the core is ferromagnetic with nuclear density. A solid foundation of the model for magnetized quark nuggets (\mqns) and predictions were developed over the years in for example \cite{VanDevender2020a, VanDevender2020, VanDevender2021, Sloan2021, VanDevender2021a, VanDevender2024}. If \mqns{} exist, interesting implications on features for galaxy clusters and observations would follow from their properties. First, \mqns{} would propose an additional source for magnetic fields in galaxy clusters as multiple \mqns{} could possibly form -- when aligned -- a large magnetic field, and \mqns{} as additional magnetic field sources are not far-fetched. Second, if \mqns{} propagated through the ICM, electrons will be accelerated by their strong magnetic fields due to an additional Lorentz force component. Since \mqns{} could be aligned to form large-scale magnetic fields, they could further act as a Faraday rotating medium of polarised synchrotron emission from the ICM. Since no studies were found addressing these large-scale implications of \mqns, these features are purely speculative, and it is important to verify if these implications hold true or not by analyzing simulations and observations.

The physical effect of magnetisation in quark composites serving as dark matter is that their effective radius would be influenced by the surrounding magnetic field. Furthermore, studies show that the strong magnetic field in \mqns{} influences the interaction with the surrounding plasma as each \mqn{} would develop a magnetopause that causes them to lose kinetic energy while moving through the plasma \citep{VanDevender2017, VanDevender2020a}. This would have the effect that dark matter halos of substructures in galaxy clusters would experience a frictional component. By estimating an on-the-fly stopping power effect for a dark matter halo, it would be possible to observe a change in the radial dark matter profiles. On the other hand, the aforementioned studies focused on magnetized quark nuggets with a different composition in the quark core and slightly different properties than axion quark nuggets. \cite{Santillan2020} studied magnetic effects specifically for quark nuggets that contain an axion domain wall in their structure and suggested that once an axion domain wall is present, a ferromagnetic state cannot be established. They concluded, however, that this state might be possible for ordinary matter nuggets.

In conclusion, the general theory of \aqns{} and similar derivations of this model yield physically interesting outcomes. Remarkably, the \aqn{} model was initially introduced as a consequence of a solution for the \cp{} problem. Long-lived cosmological problems are in return naturally resolved by this model. Observable features propose predictions that can be validated by numerical methods as we showed throughout this study. This work serves as a first foundation to show that observable \aqn{} features can be studied in cluster environments using cosmological simulations. Our studies show that it might be possible to detect \aqn{} signatures, especially in galaxy clusters with low synchrotron emission from the ICM. Our results also show that to disentangle pure \aqn{} contributions from the radiation background in galaxy clusters, we need improved strategies and more sophisticated models to predict the non-thermal emission of galaxy clusters.

\begin{acknowledgements}
We thank the Center for Advanced Studies (CAS) of LMU Munich for hosting the collaborators of the LOCALIZATION project for a week-long workshop. This work was supported by the grant agreements ANR-21-CE31-0019 / 490702358 from the French Agence Nationale de la Recherche / DFG for the LOCALIZATION project. KD acknowledges support by the Excellence Cluster ORIGINS which is funded by the Deutsche Forschungsgemeinschaft (DFG, German Research Foundation) under Germany’s Excellence  Strategy – EXC-2094 – 390783311 and funding for the COMPLEX project from the European Research Council (ERC) under the European Union’s Horizon 2020 research and innovation program grant agreement ERC-2019-AdG 882679. XL, LVW, AZ and FM acknowledge the support from NSERC. The calculations for the hydro-dynamical simulations were carried out at the Leibniz Supercomputer Center (LRZ) under the project pn68na.  
\end{acknowledgements}

\clearpage


\bibliographystyle{aa} 
\bibliography{literature.bib} 


\begin{appendix} 

\section{Supplementary Figures}

\begin{figure}[h!]
    \centering
    \includegraphics[width=1.0\linewidth]{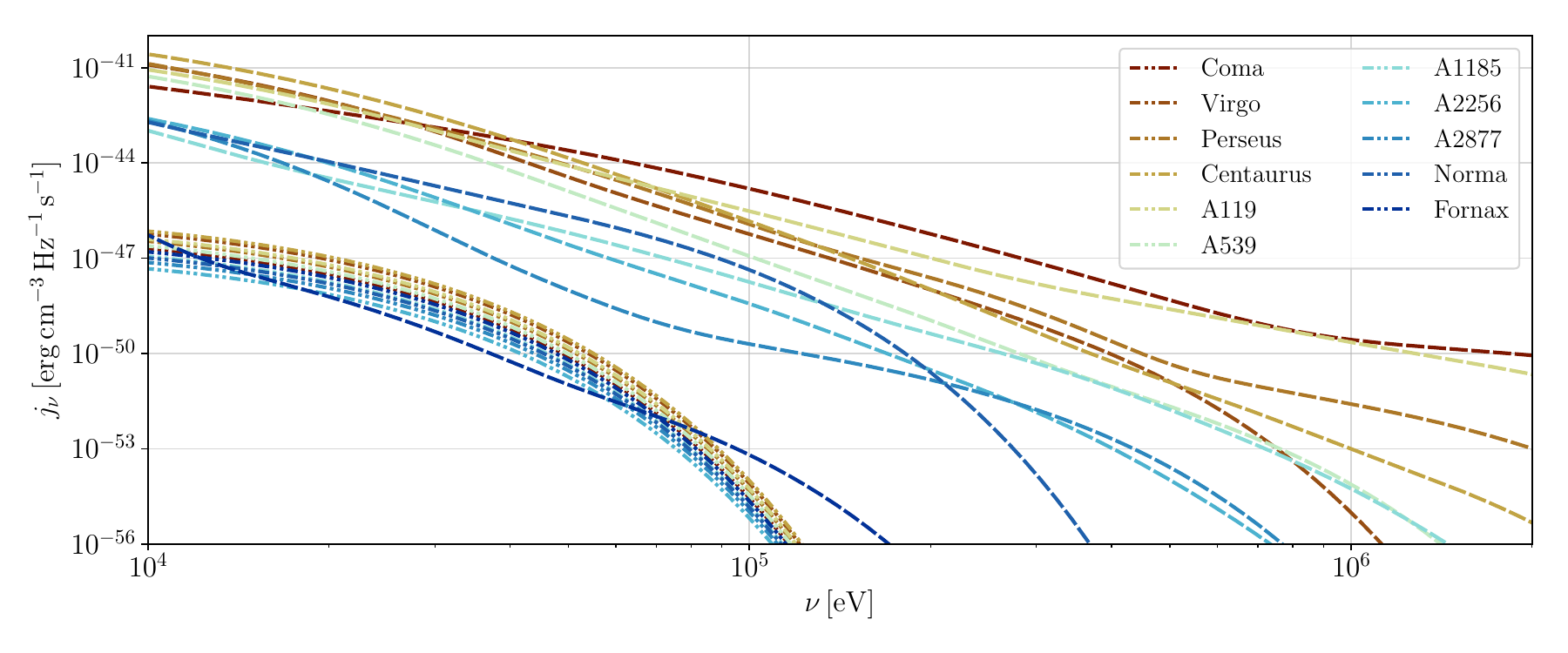}
    \caption{Spectral plot for non-thermal \aqn{} emission and thermal Bremsstrahlung emission of the hot ICM incorporating the spectra from \sampleb. Fornax, being the only galaxy cluster, shows a small region where \aqn{} emission dominates gas emission.}
    \label{fig:high_energy_spec}
\end{figure}

In \autoref{fig:high_energy_spec} we show the high energy part of the spectra for 11 cross-identified clusters from our sample. The different amounts of the contribution of the AQN emission to the X-ray signal can be clearly seen for the individual clusters. For Fornax, the AQN signal in our simulation even over-shines the thermal emission in the hard X-ray regime.

\begin{figure}[bh!]
    \centering
    \includegraphics[width=1.0\linewidth]{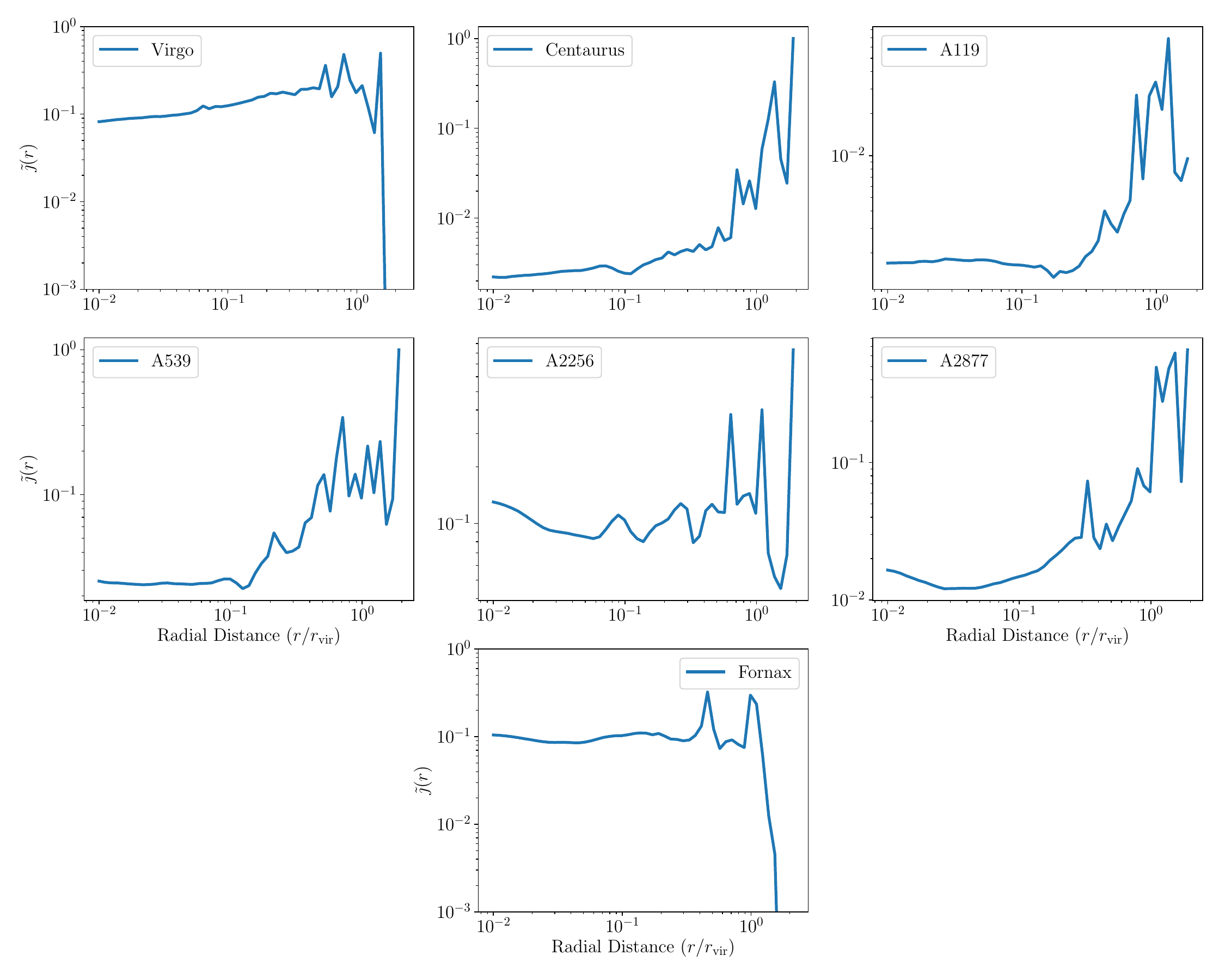}
    \caption{Radial profiles of $\specratio(r)$ (see \autoref{eq:all_specratio}) for cross-correlated galaxy clusters that showed a thermal \aqn{} emission excess in the spectral features. Galaxy clusters with high $\specratio$ in \autoref{fig:intspecratio} show higher $\specratio(r)$ in central regions of the galaxy cluster.}
    \label{fig:emradprof}
\end{figure}

In \autoref{fig:emradprof} we show the radial profiles of $\specratio(r)$ (see \autoref{eq:all_specratio}) for 7 of the cross-identified clusters. They are the same that showed a thermal \aqn{} emission excess in the spectral features. Note that for some of the systems (especially Virgo, Fornax and A2256) the contribution within the core of the clusters by the AQNs emission is reaching a 10\% level.

\end{appendix}

\end{document}